\documentclass[aps,pra,showpacs,twocolumn,floatfix]{revtex4}
\usepackage{graphicx}
\usepackage{dcolumn}
\usepackage{amsmath}

\begin{document}

\title {Relativistic many-body calculations of energies for
core-excited states in sodiumlike ions}

\author{U.I. Safronova}
\email{usafrono@nd.edu}
\author{W. R. Johnson}
\email{johnson@nd.edu}
\homepage{www.nd.edu/~johnson}
\author{M.S. Safronova}
 \altaffiliation[Current address: ]{Electron and Optical Physics
Division, National Institute of Standards and Technology,
Gaithersburg, MD, 20899-8410}
\affiliation{%
Department of Physics, 225 Nieuwland Science Hall\\
University of Notre Dame, Notre Dame, IN 46566}%

\author{J. R. Albritton}
\affiliation{Lawrence Livermore National Laboratory, PO Box 808,
Livermore, CA 94551}

\date{\today}

\begin{abstract}

Energies of $(2s^22p^53l3l')$ and  $(2s2p^63l3l')$ states for
sodiumlike ions with $Z$ =14--100 are evaluated to second order in
relativistic many-body perturbation theory starting from a
neonlike  Dirac-Fock potential. Second-order Coulomb and
Breit interactions are included. Correction for the
frequency-dependence of the  Breit interaction is taken into
account in lowest order. The Lamb shift correction to energies
is also included in lowest order.  Intrinsic particle-particle-hole
contributions to  energies are found to be 20-30\% of the
 sum of  one-
 and two-body contributions.  Comparisons are
made with available experimental data.
 These calculations are presented as a
theoretical benchmark for comparison with experiment and theory.
\end{abstract}



\maketitle

\section{Introduction}

Core-excited states of Na-like ions have
two valence electrons outside and one hole inside a closed $n$=2
core and, therefore, present an excellent model for studying
strong correlations in systems with closely-spaced  levels. We
carry out second-order relativistic many-body perturbation theory
(RMBPT) calculations for Na-like ions starting from a
$1s^22s^22p^6$ Dirac-Fock (DF) potential. All possible $2l$ holes
and $3l'3l''$ particles leading to 121 odd-parity and 116
even-parity $3l'3l''[J_{1}]2l^{-1}(J)$ states are considered. We
calculate energies of the 237 core-excited states  together with
the $3l$ singly-excited states in Na-like ions with nuclear
charges $Z$ ranging from 14 to 100.

 Many experimental values for the energy levels and
fine-structure intervals are now available up to very high
nuclear charge ($Z$=57) for $3l'3l''[J_{1}]2l^{-1}(J)$ levels;
additionally, experimental rates for some transitions between
these levels are available.
 The objective of this paper is to present a comprehensive set of
calculations for $3l'3l''[J_{1}]2l^{-1}(J)$ energies  to compare
with previous calculations  and experiments for the entire Na
isoelectronic sequence. Most earlier measurements and calculations
focused on the $3s^2[^1S]2p^{-1}\ ^2P_J$ states and low-lying
$3s3p[^3P]2p^{-1}\ ^4D_J$ and $3s3d[^3D]2p^{-1}\ ^4F_J$ levels.
Very few results exist for other $3p^22p^{-1}$, $3s3p2p^{-1}$, and
$3s3d2p^{-1}$ levels. The large number of possible transitions
have made experimental identification difficult. Experimental
verification should become simpler and more reliable using this
more accurate and complete set of calculations.

We start with a brief overview of previous theoretical and
experimental studies of properties of core-excited states
in sodium-like ions.
 Energies, collision strengths, and oscillator strengths for
$2s^22p^63l-2s^22p^53l^{\prime }3l^{\prime \prime }$ and
$2s^22p^63l-2s2p^63l^{\prime}3l^{\prime \prime }$ 
transitions  in Na-like ions with nuclear charges $22 \leq Z \leq 62$
were calculated by \citet{m1}, where
energies and mixing coefficients were obtained using the Cowan
code \cite{cowan} with a scaling factor $(Z-3.9)/Z$. In the
paper by \citet{m2}, Auger and radiative transition energies
and rates were calculated for 18 ions with nuclear charges in the
range  $18 \leq Z\leq 92$ using the multiconfiguration Dirac-Fock (MCDF)
method.
Wavelengths, radiative transition rates, Auger rates,
satellite intensity
factors, and fluorescence yields were presented by \citet{m3} for
dielectronic
satellites of 14 neon-like ions (Ar$^{8+}$ - W$^{64+}$)
including $2s^22p^63l$--$2s^22p^53l^{\prime }3l^{\prime \prime}$ and
$2s^22p^63l$ -$2s2p^63l^{\prime }3l^{\prime \prime }$ transitions.
The calculations in \cite{m3} were made using the {\sc yoda} code
which is based on multiconfiguration
relativistic functions for the bound states and
distorted-wave Dirac wave functions for continuum electrons.
New comprehensive theoretical analyses
of the  $2s^22p^53lnl'$ and $2s2p^63lnl'$ ($n$=3-8) doubly-excited
states in Fe$^{15+}$ and Cu$^{18+}$ were presented
by \citet{bruch,safr}. In these papers, energy levels, radiative
transitions probabilities, and autoionization rates were
calculated by using the Cowan and {\sc yoda} codes.  Recently, R-matrix
calculations of electron-impact collision strengths for excitation
from the inner $L$-shell into doubly excited states of Fe$^{15+}$
were presented by \citet{baut}. Energy positions of
the $2s^22p^63l$ and $2s^22p^53l3l'$ states were calculated in
that work using the {\sc superstructure} code of \citet{eiss74}. It was
shown in \cite{baut} that disagreement between calculations
and data recommended by \citet{sugar} and
\citet{shirai} ranges from 0.5\% to 5\%.

Studies of the $2s^22p^53lnl^{\prime }$ configurations in Na-like ions
are of continuing interest from both  theoretical and
experimental  points of view. Experimentally, these configurations
 are studied by
photon and electron emission spectroscopy. To our knowledge, the
first measurements of $2p-3s$ and $2p-3d$ transitions in Na-like Fe
(which included the $2s^22p^53l3l^{\prime }$
configurations) using photon emission spectroscopy were done by
\citet{a1} in \citeyear{a1}. The soft x-ray spectra
produced by a focused laser source and a vacuum spark device were
obtained in \cite{a1} using a grazing incidence
spectrograph. Identification of the spectra was based on
theoretical calculations using the Cowan code. The $2s^22p^53s3p$
$^4D_{7/2}$ - $2s^22p^53s3d$ $^4G_{9/2}$ transition in FeXVI was
observed by the beam-foil technique by \citet{a2}.
The identification of this transition was based on a study
 of the difference in wave numbers
between experimental values and theoretical predictions obtained from
the Cowan code in Na-like ions from S~VI to Cu~XIX.
 Auger electron spectra of Na-like Fe ions excited in
collisions of 170-keV Fe XVIII on He and Ne were studied
by \citet{a3}. The dominant spectral
structures were due to Auger decay of states with
$2s^22p^53l3l^{\prime}$ configurations in FeXVI. Special attention was
paid to the metastable level $2s^22p^53s3p$ $^4D_{7/2}$ which was
studied in detail; the absolute energy of this level was measured. Three
of the 17 lines observed in this experiment were
identified with states from $2s^22p^53l4l^{\prime }$
configurations. This analysis was based on theoretical predictions using
the multiconfigurational Dirac-Fock method including the transverse Breit
interaction. A few years later this analysis was extended to include an
isoelectronic study of double electron capture in slow ion-atom
collisions of F-like ions (Si$^{5+}$, Ar$^{9+}$, Sc$^{12+}$,
Ti$^{13+}$, Fe$^{17+}$, and Cu$^{20+}$) with He atoms \cite{m4}.
It was reported that the metastable level $2s^22p^53s3p$
$^4D_{7/2}$ was observed in all Auger spectra except Si$^{3+}$.
Recently in \cite{a5}, a study of the double electron capture in
low-energy Fe$^{17+}$ + He collisions was conducted.  Except for
the confirmation of the identification of the 12 peaks in the
225 - 340~eV interval, predictions were made for identification of four
additional peaks in the 585 - 670~eV interval. A comparison of
solar and tokamak FeXVII x-ray spectra with synthetic spectra
including dielectronic satellites with
$2s^22p^6nl-2s^22p^53snl$ and $2s^22p^6nl-2s^22p^53dnl$ transitions was
presented by \citet{phil}. Atomic
data were calculated in \citet{phil} using the {\sc superstructure} code.
Recently, the three Fe$^{15+}$ inner-shell
satellite lines were thoroughly studied at the Lawrence Livermore
National Laboratory electron beam ion trap {\sc ebit-ii} by  \citet{brown}.

In the present paper, we use RMBPT to determine energies of $2s^22p^53l3l'$ and
$2s2p^63l3l'$ states for Na-like ions with nuclear charges in the
range of $Z$ = 14--100. Our calculations are carried out to
second order in perturbation theory and include second-order
Coulomb and Breit  interactions. Corrections for the
frequency-dependent Breit interaction are taken into account in
the lowest order. Screened self-energy and vacuum polarization
data given by Blundell \cite{qed} are used to determine the QED
correction.

Our perturbation theory calculations are carried out using
single-particle orbitals calculated in the DF potential of the
Ne-like core. As a first step, we determine and
store the single-particle contributions to the energies for three
$n$=2 hole states ($2s$, $2p_{1/2}$, and $2p_{3/2}$) and   the
five $n$=3 valence states ($3s$, $3p_{1/2}$, $3p_{3/2}$,
$3d_{3/2}$, and $3d_{5/2}$) in lowest, first, and second orders.
 Next, we
evaluate and store the 155 two-particle  $\langle 3l 3l'\: J |
H^{\rm eff} |3l''3l''' \: J \rangle$ matrix elements  and the 178
hole-particle  $\langle 2l 3l'\: J | H^{\rm eff} |2l''3l''' \: J
\rangle$ matrix elements of the effective Hamiltonian in the
first and second orders. It should be noted that these one-particle,
two-particle, and hole-particle matrix elements were used
previously to evaluate energies of the $3l3l'$ levels in
Mg-like ions \cite{mg} and energies of the $2l^{-1}3l'$
levels in neonlike ions \cite{neon}.  Finally, second-order
particle-particle-hole matrix elements are evaluated. Combining
these data using the method described below, we calculate one-,
two-, and three-body contributions to the energies of Na-like
ions.

The present calculations are compared with experimental and
predicted results from Refs.~\cite{a2,a5,b86,b95}.

\section {Method}
The RMBPT formalism developed previously \cite{bor2,bor3,alum} for
B-like and Al-like ions is used here to describe the perturbed wave
functions and to obtain the second-order energies \cite{bor2} of the
core-excited states of Na-like ions. The particle-particle-hole
contribution to the energies of  those states is, however, different from the
 three-particle contribution to the energies of
  B-like and Al-like ions, and  is derived in the Appendix. Including
hole-particle matrix elements in Na-like ions leads to 237
$3l3l'2l^{-1}$ states instead of 148 $3l3l'3l''$ states of Al-like
ions  or 15 $2l2l'2l''$ of B-like ions and, consequently, to more laborious
numerical calculations.

\subsection{Model space}
The model space for core-excited $3l3l'2l^{-1}$  states of
Na-like ions includes 121 odd-parity states consisting of 25
$J$=1/2 states, 37 $J$=3/2 states, 31 $J$=5/2 states, 19 $J$=7/2
states, 7 $J$=9/2 states, and two $J$=11/2 states. Additionally,
there are 116 even-parity states consisting of  25 $J$=1/2 states,
35 $J$=3/2 states, 31 $J$=5/2 states, 17 $J$=7/2 states,  7
$J$=9/2 states, and one $J$=11/2 state.

The evaluation of the second-order energies for the
$3l3l'[J_{1}]2l^{-1} (J)$ states in Na-like ions follows the pattern
of the corresponding calculation for Mg-like and Ne-like ions
given in Refs.~\cite{mg,neon}. In particular, we use the
second-order one-particle and two-particle matrix elements for Mg-like
ions calculated in \cite{mg} and hole-particle matrix elements for
Ne-like ions calculated in \cite{neon}. These matrix elements
are recoupled as described
below to obtain the one-body and two-body contributions for
Na-like ions.
 We will discuss how these matrix
elements are combined to obtain the one-body and two-body
contributions to energies of Na-like ions. We refer the reader to
Refs.~\cite{mg,neon} for a discussion of the evaluation of
the one-body and
two-body matrix elements. An intrinsic
particle-particle-hole diagram also contributes to the
second-order energy for Na-like ions. The
expression for particle-particle-hole diagram differs from the
expression for the  three-particle diagram given in
Ref.~\cite{bor2} owing  to different angular parts.
The expression for the intrinsic particle-particle-hole
diagram is derived in Appendix.

\subsection{Energy-matrix elements }

\begin{figure*}
\centerline{\includegraphics[scale=0.35]{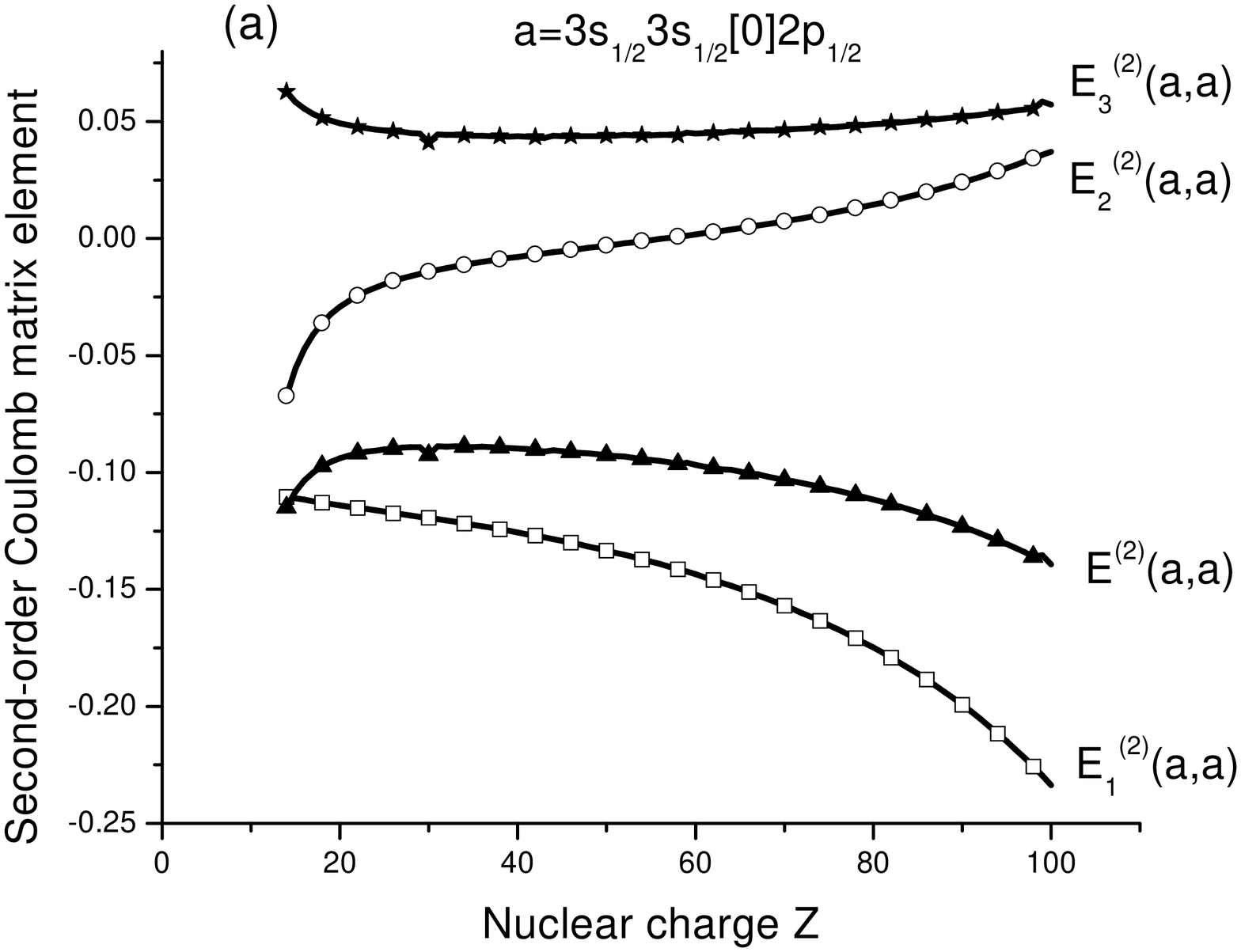}
             \includegraphics[scale=0.35]{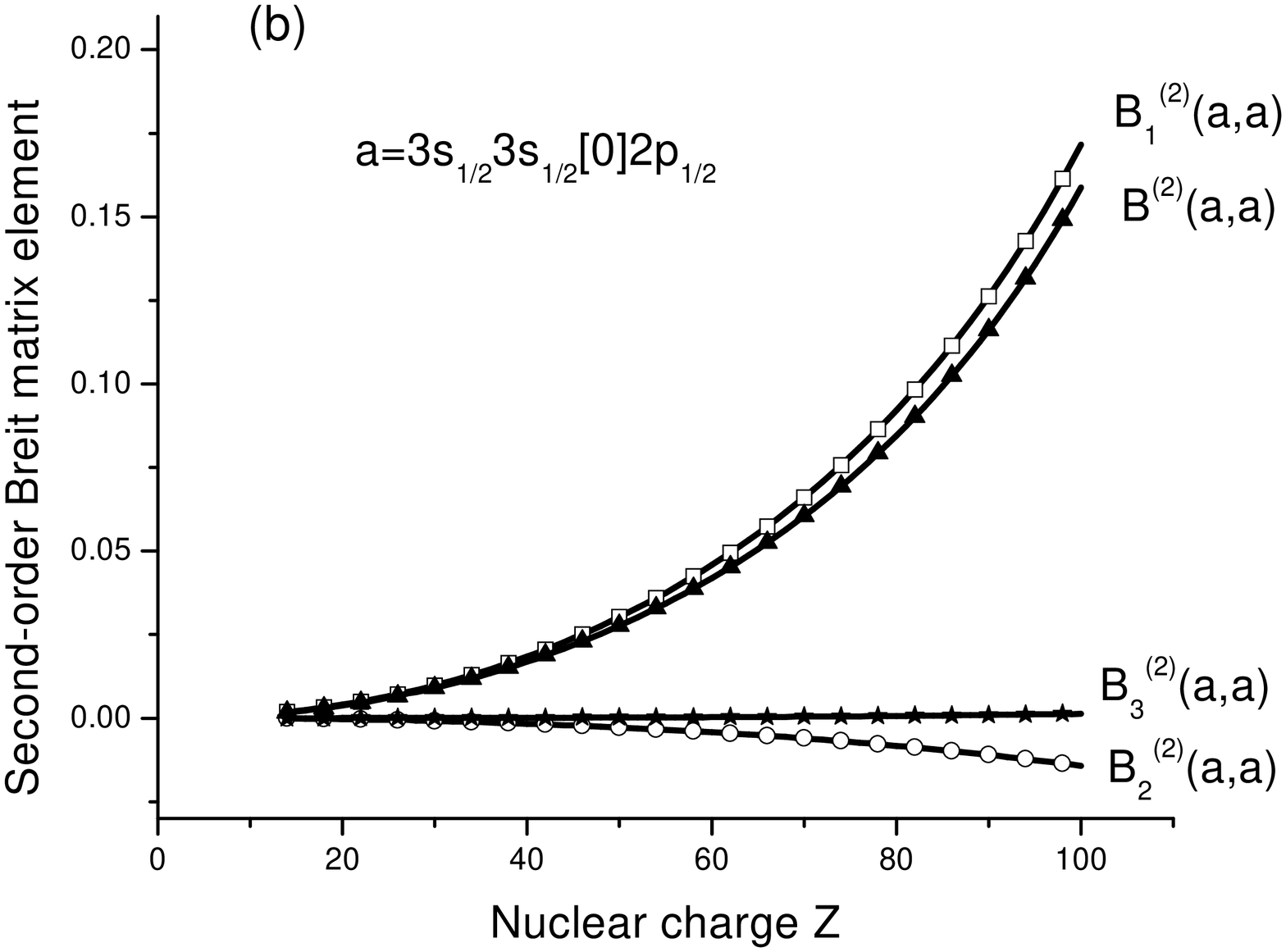}}
\caption{Second-order Coulomb (a) and Breit (b) energy
contributions to diagonal matrix elements in Na-like ions.}
\label{e2-d}
\end{figure*}
\begin{figure*}
\centerline{\includegraphics[scale=0.35]{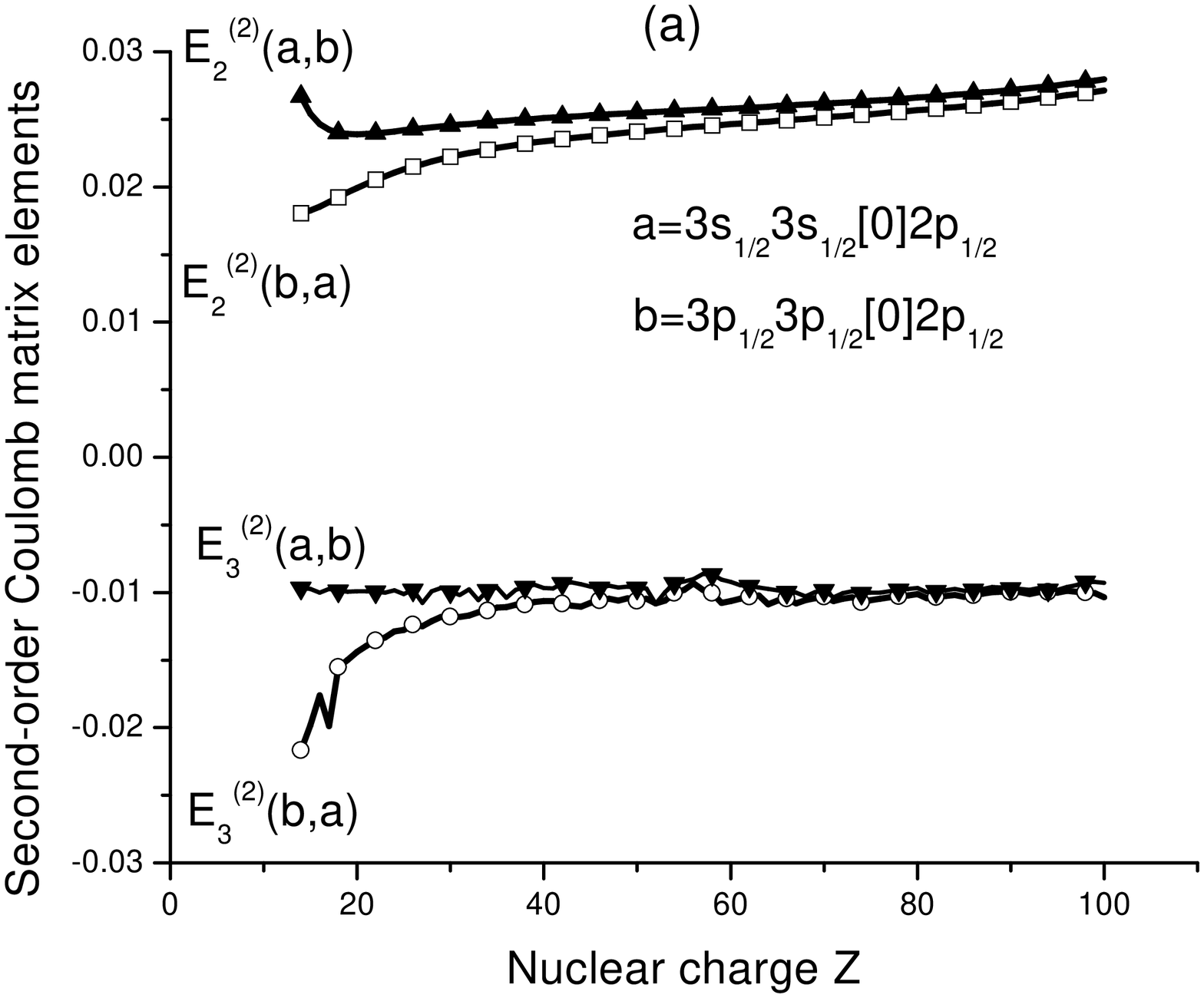}
             \includegraphics[scale=0.35]{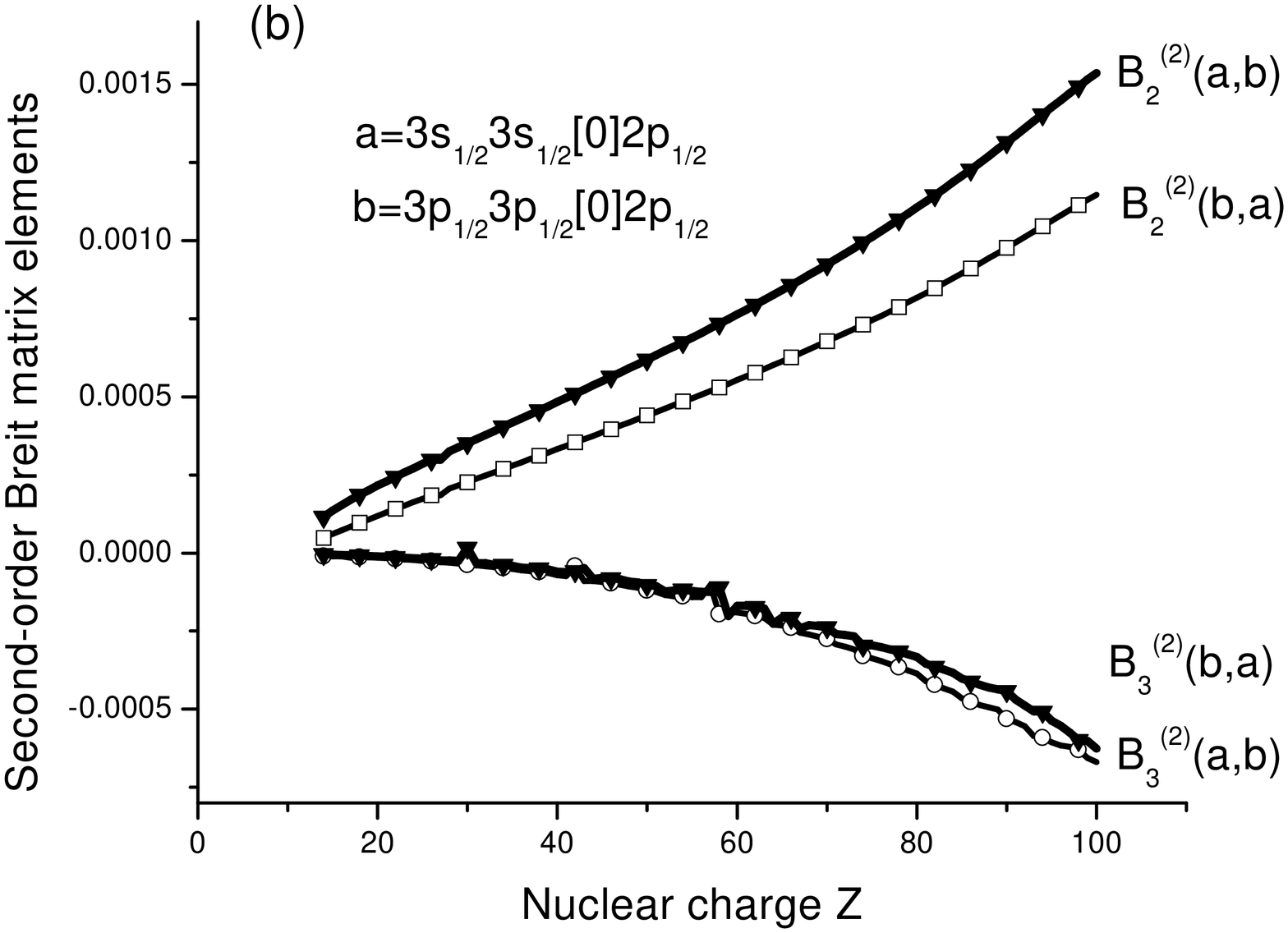}}
\caption{Second-order Coulomb (a) and Breit (b) energy
contributions to non-diagonal matrix elements in Na-like ions.}
\label{e2-nd}
\end{figure*}
In Fig.~\ref{e2-d}a, we present the one-body,
$E_{1}^{(2)}(a,a)$, two-body, $E_{2}^{(2)}(a,a)$, and three-body,
$E_{3}^{(2)}(a,a)$ Coulomb contributions to the diagonal matrix element
with $a=3s_{1/2}3s_{1/2}[0]2p_{1/2}\ (1/2)$
as functions of $Z$. Here and in the sequel, we drop the superscript
$-1$ on the $n=2$ hole state for simplicity. 
We see from Fig.~\ref{e2-d}a that the
value of $E_{3}^{(2)}(a,a)$ is almost constant for all $Z$. The
one-body contribution is the largest one.
The sum of  $E_{2}^{(2)}(a,a)$ and  $E_{3}^{(2)}(a,a)$ values  is almost
equal to zero for small-$Z$ ions; however, the
three-body contribution becomes more important
with increasing $Z$. As a result, the total second-order Coulomb contribution
$E^{(2)}(a,a)$=$E_{1}^{(2)}(a,a)$+$E_{2}^{(2)}(a,a)$+$E_{3}^{(2)}(a,a)$
differs from the one-body contribution and  is almost constant for
all $Z$.
In Fig.~\ref{e2-d}b, we present the one-body,
$B_{1}^{(2)}(a,a)$, two-body, $B_{2}^{(2)}(a,a)$, and three-body,
$B_{3}^{(2)}(a,a)$ Breit-Coulomb contributions to the diagonal matrix element
with $a=3s_{1/2}3s_{1/2}[0]2p_{1/2}\ (1/2)$
 as function of $Z$.
 The one-body
 contribution is the largest among other contributions and the
total Breit second-order contribution
$B^{(2)}(a,a)$=$B_{1}^{(2)}(a,a)$+$B_{2}^{(2)}(a,a)$+$B_{3}^{(2)}(a,a)$
is almost equal to the one-body contribution.
The one-body contribution vanishes for non-diagonal
matrix elements. 

In Fig.~\ref{e2-nd}a, we illustrate the contributions
$E_{2}^{(2)}$ and  $E_{3}^{(2)}$ for  two
non-diagonal elements $E_{i}^{(2)}(a,b)$ and $E_{i}^{(2)}(b,a)$
with $a$=$3s_{1/2}3s_{1/2}[0]2p_{1/2}(1/2)$ and
$b$=$3p_{1/2}3p_{1/2}[0]2p_{1/2}(1/2)$. The corresponding
Breit contributions are plotted in Fig.~\ref{e2-nd}b. We  see from
Fig.~\ref{e2-nd}a  that the ratio of $E_{3}^{(2)}$  to
$E_{2}^{(2)}$ is about 30--50\% for high-$Z$ ions.
Fig.~\ref{e2-nd}a illustrates the asymmetry of the nondiagonal
energy matrix elements for low-$Z$ ions. The asymmetry of the
two-particle second-order matrix elements in RMBPT calculation was
discussed in Ref.~\cite{be3}. Values of $E_{2}^{(2)}$ and
$E_{3}^{(2)}$ have different signs and the total second-order
contribution $E^{(2)}$ is about half of  $E_{2}^{(2)}$.
The behavior of non-diagonal second-order Breit
contribution is shown in Fig.~\ref{e2-nd}b.
The $B_{2}^{(2)}$ and $B_{3}^{(2)}$
increase with $Z$; however, they have opposite signs.
The total second-order contribution $B^{(2)}$ can be
about half of  $B_{2}^{(2)}$. Finally, we observe
some small deviations from smooth $Z$-dependencies of $E^{(2)}_3$
and $B^{(2)}_3$ in Figs.~\ref{e2-nd}a and
\ref{e2-nd}b.  As we can see from Eq.~(\ref{eq10}) in the Appendix, 
the denominator
${\varepsilon (v)+\varepsilon (w)-\varepsilon (a)-\varepsilon (n)}$
in the expression for the three-body diagram
contains  two valence energies ${\varepsilon (v)}$,
${\varepsilon (w)}$, one hole energy ${\varepsilon (a)}$, and
one virtual state energy ${\varepsilon (n)}$.
 The sum ${\varepsilon (3s)+\varepsilon
(3s)-\varepsilon (2p)}$ is positive; therefore, the denominator
 can be close to zero when ${\varepsilon
(n)}$ is positive. In such a case, a
singularity  arises from the contribution of the continuous part
of spectra to the sum over states for the $E_{3}^{(2)}$ and
$B_{3}^{(2)}$ matrix elements, resulting in sharp
features in the corresponding $Z$-dependencies.

\subsection{Eigenvalues
and eigenvectors for core-excited states}
After evaluating the energy matrices,  we calculate eigenvalues
and eigenvectors for states with given values of $J$ and parity.
There are two possible methods  to  carry out the diagonalization:
(a) diagonalize the sum of  zeroth-  and  first-order  matrices,
then calculate the second-order contributions using the resulting
eigenvectors or (b) diagonalize the  sum of  the zeroth-,  first-,
and  second-order matrices together. Following Ref.~\cite{bor2},
we choose the second method here.

\begin{figure*}[tbp]
\centerline{\includegraphics[scale=0.35]{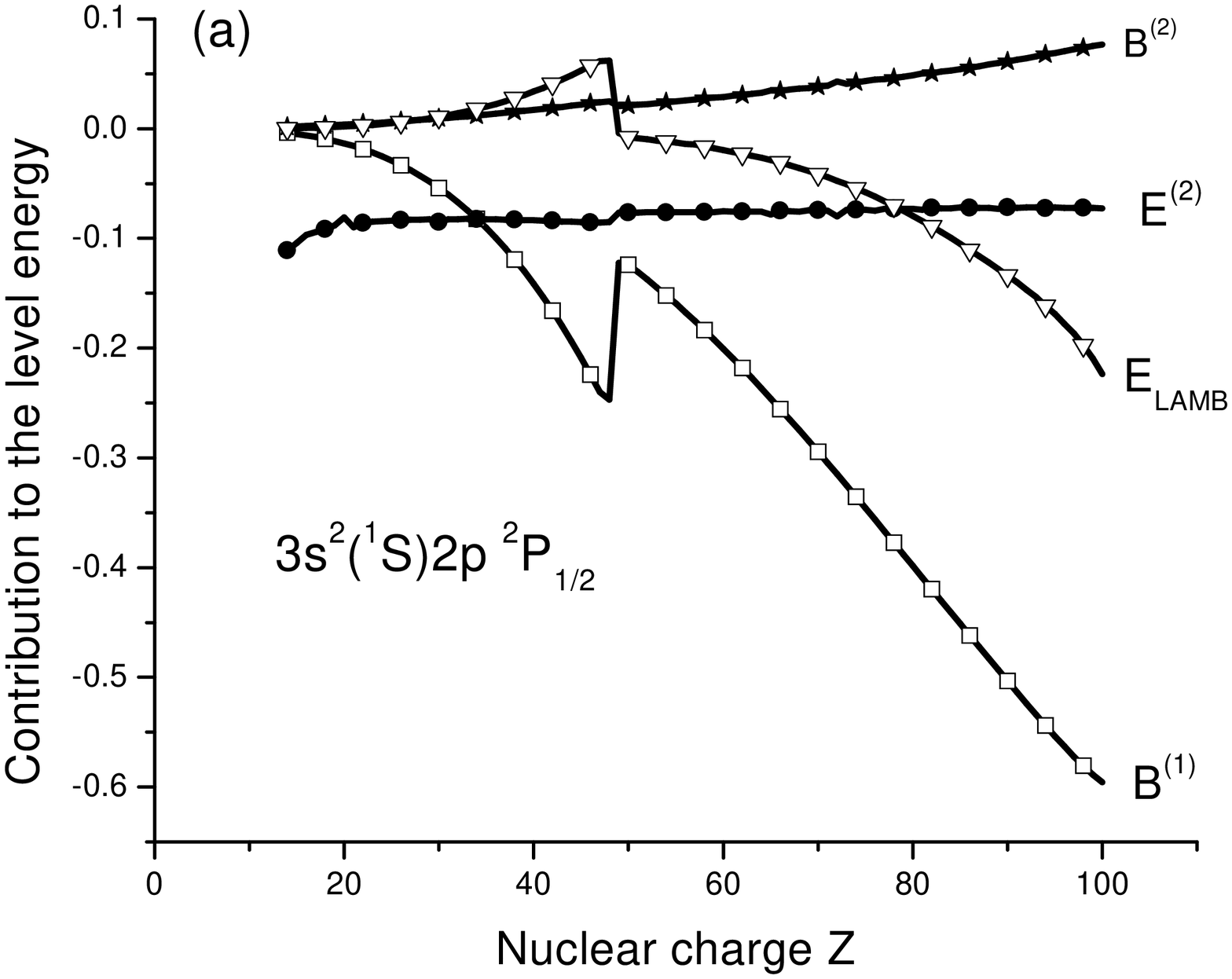}
            \includegraphics[scale=0.35]{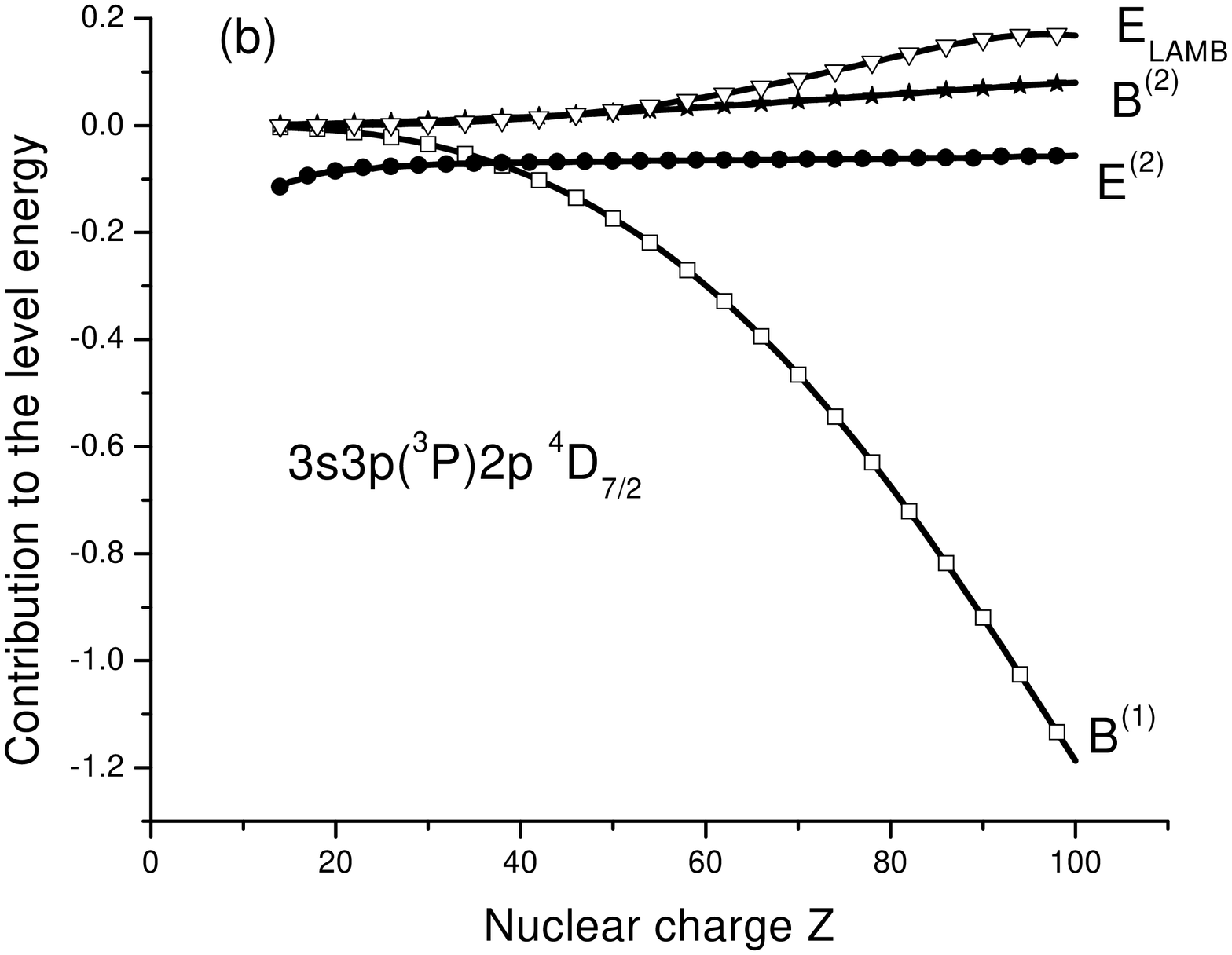}}
\caption{Contributions to the energies of the $3s^2(^1S)2p\
^2P_{1/2}$ (a) and $3s3p(^3P)2p\ ^4D_{7/2}$  (b) levels of Na-like
ions.} \label{contr}
\end{figure*}
The importance of  second-order contributions to the energies
is illustrated in Fig.~\ref{contr}. In this figure, the
second-order energy $E^{(2)}$, the
first- and second-order Breit contributions $B^{(1)}$ and $B^{(2)}$, and
the QED contribution $E_{\rm {LAMB}}$ are plotted as functions of
nuclear charge $Z$ for the
$3s^2(^1S)2p\ ^2P_{1/2}$ (a) and $3s3p(^3P)2p\ ^4D_{7/2}$ (b)  states of
Na-like ions. We can see from the figure that $E^{(2)}$
dominates up to $Z$=34 and $Z$=37 for the  $3s^22p\
^2P_{1/2}$ and $3s3p(^3P)2p\ ^4D_{7/2}$ states, respectively. The QED
contribution $E_{\rm {LAMB}}$ is smaller than the $E^{(2)}$ energy
 up to $Z$=50 for the energy level shown in Fig.~\ref{contr}a
  and $Z$=60 for the energy level shown in Fig.~\ref{contr}b.  The
second-order   Breit energy  $B^{(2)}$
 is  smaller than the
second-order energy $E^{(2)}$ up to $Z$=97 (Fig.~\ref{contr}a) and
$Z$=83 (Fig.~\ref{contr}b), respectively.

The sharp features in the curves  shown in the Fig.~\ref{contr}a
are explained by strong mixing of states inside
of the odd-parity complex with $J$=1/2. The main contribution to
 the $3s^2(^1S)2p\ ^2P_{1/2}$ level comes from the
$3s_{1/2}3s_{1/2}[0]2p_{1/2}$  state for  ions with small $Z$ up to
$Z$=48. Starting from $Z$=49, the main contribution to this level
comes from $3p_{1/2}3p_{3/2}[1]2p_{3/2}$ and
$3p_{1/2}3p_{3/2}[2]2p_{3/2}$ states instead of
$3s_{1/2}3s_{1/2}[0]2p_{1/2}$ state.

The particle-particle-hole levels  studied in this work may  be
divided to two groups  by their energies:
those with energies in the range of 15--23~a.u  and those
with energies greater than 25~a.u.
The first group of levels includes
 $3s_{1/2}3s_{1/2}[0]2p_j$,
$3s_{1/2}3d_j[J_{1}]2p_{j'}$,
 $3p_{j}3p_{j'}[J_1]2p_{j''}$,
and $3s_{1/2}3p_j[J_{1}]2p_{j'}$ levels (64 levels total). The
second group includes the remaining 173 levels.
The first group of levels has been studied experimentally,
while there are no experimental data for the second group.

When starting calculations from relativistic
DF wave functions,
 it is natural to use $jj$ designations for uncoupled
 energy matrix elements; however, neither $jj$ nor $LS$ coupling
describes {\em physical} states properly, except for the
single-configuration state $3p_{3/2}3d_{5/2}[4]2p_{3/2} \equiv
3p3d[^3F]2p\ ^3\! G_{11/2}$. Both designations are given in
our subsequent tables.

\begin{figure*}
\centerline{
\includegraphics[scale=0.30]{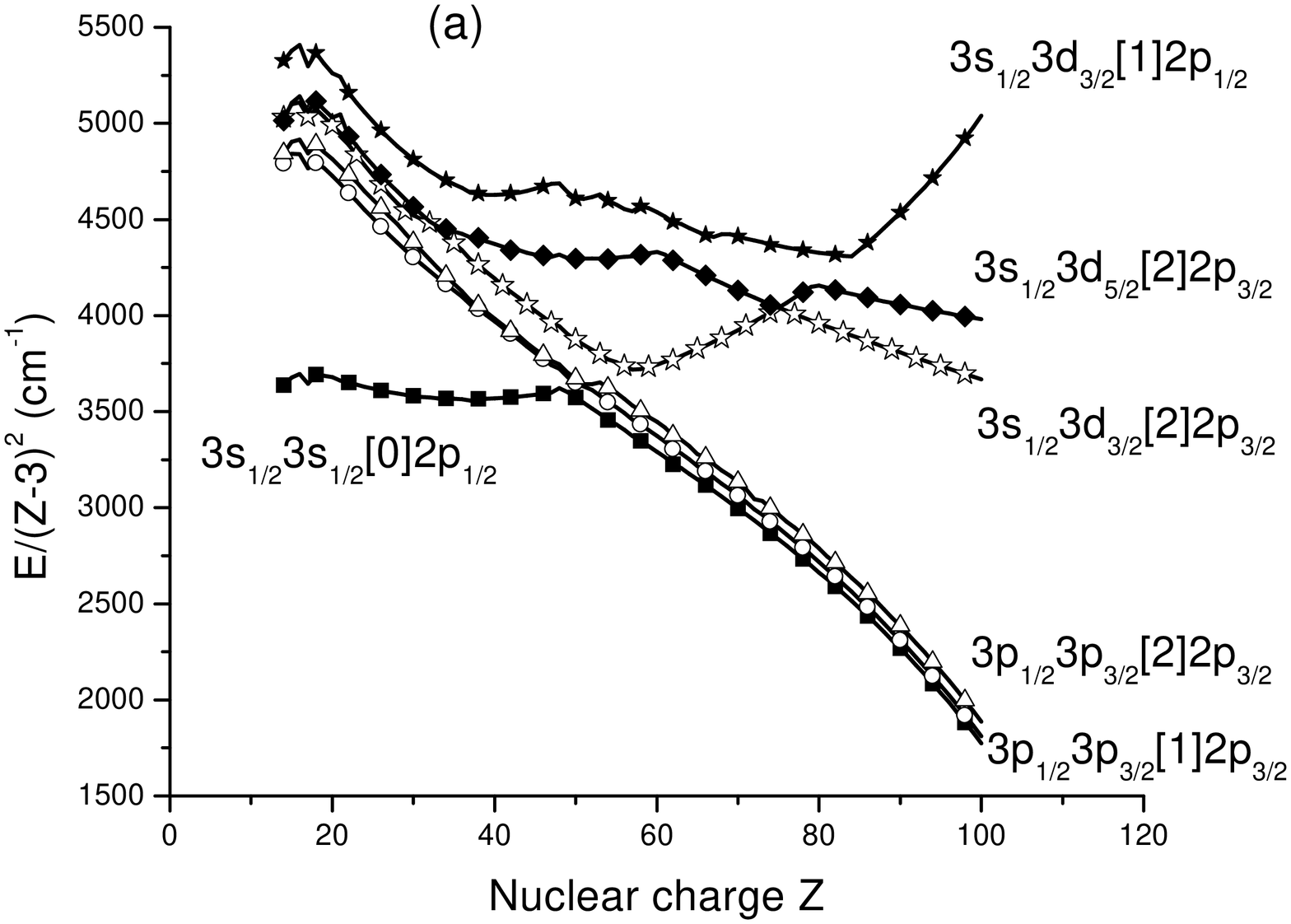}
\includegraphics[scale=0.30]{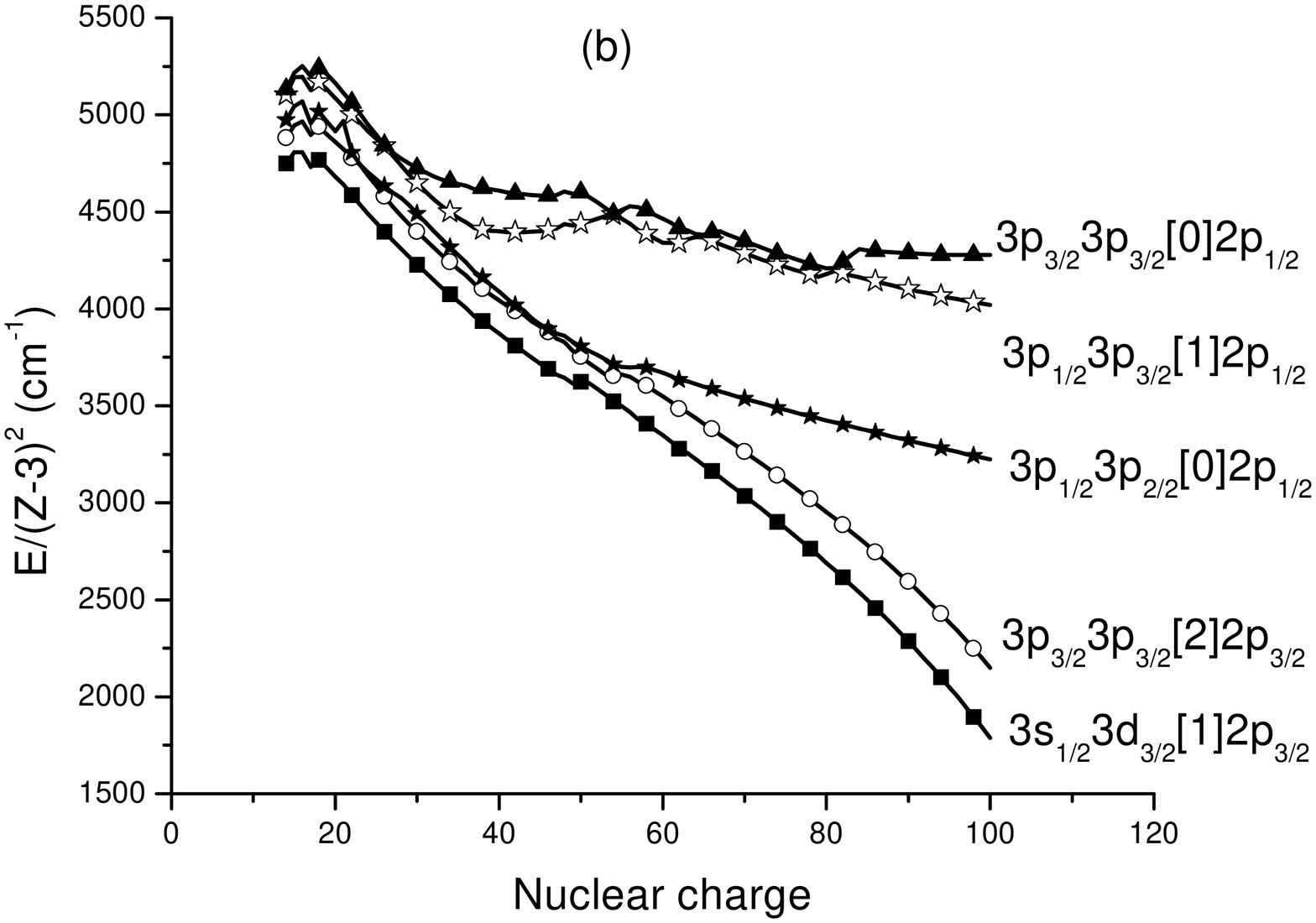}}
\centerline{
\includegraphics[scale=0.30]{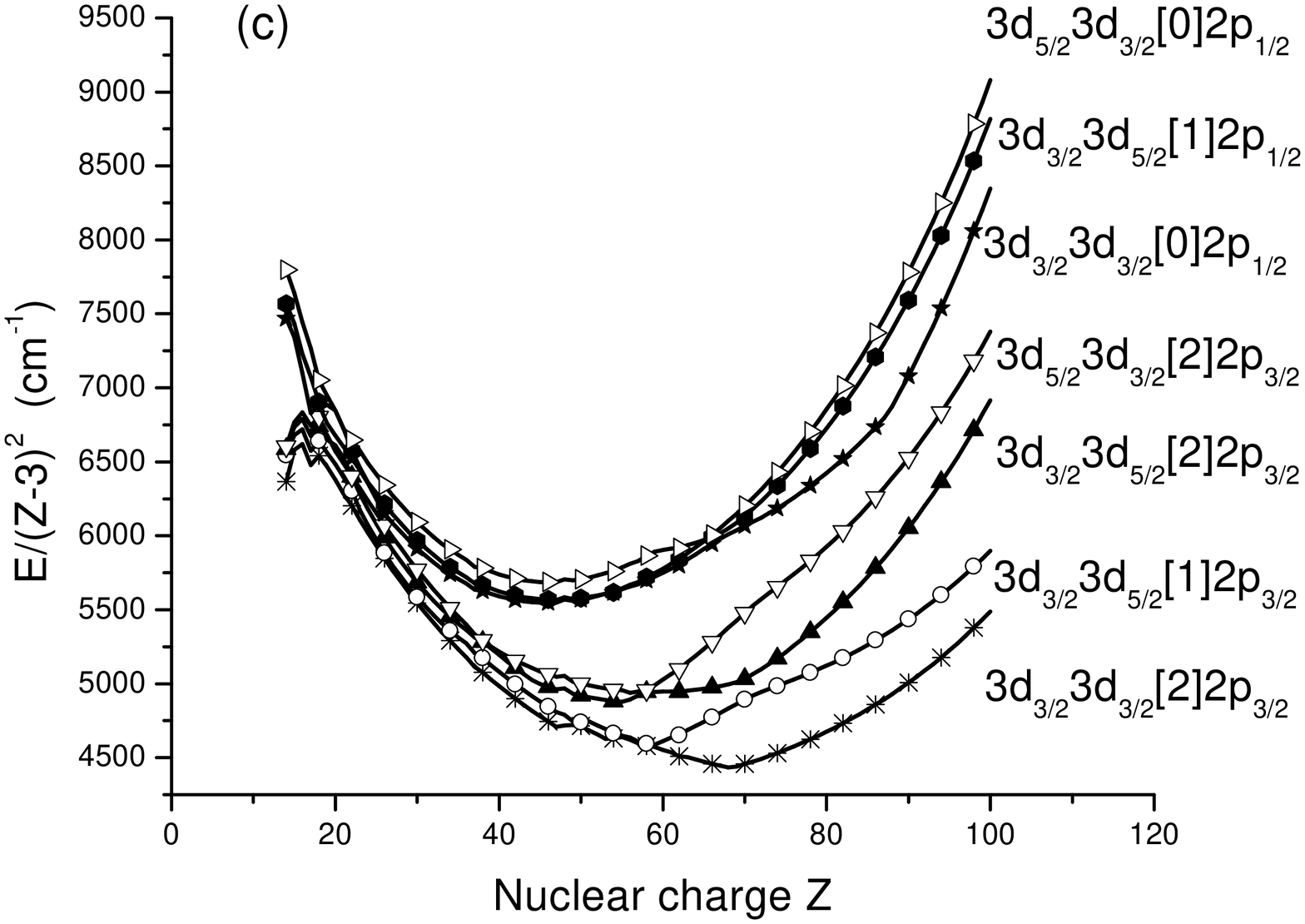}
\includegraphics[scale=0.30]{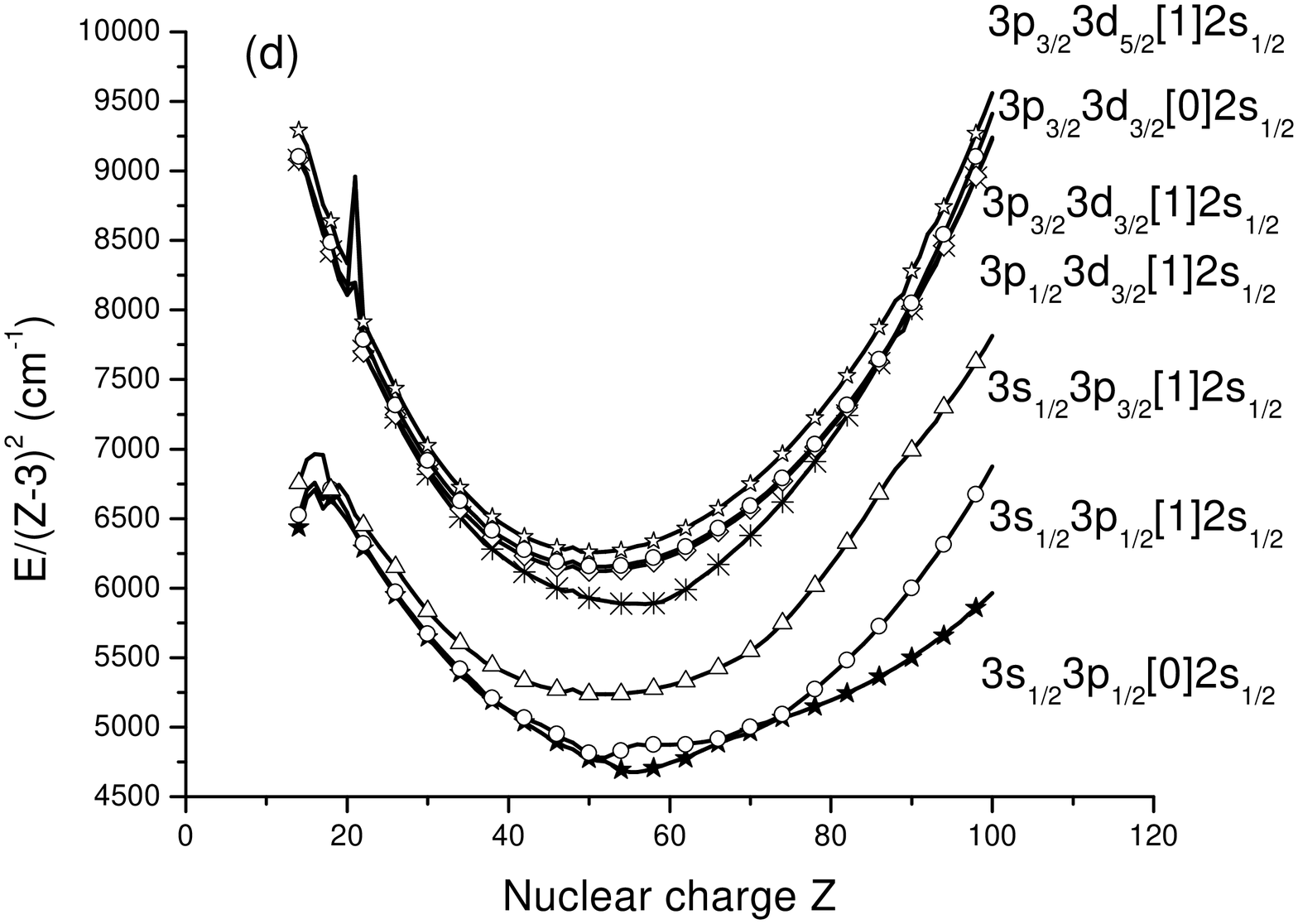}}
\caption{Energies ($E/(Z-3)^2$ in cm$^{-1}$) of odd-parity states
with $J$=1/2 as functions of $Z$.} \label{f-et}
\end{figure*}
The $Z$-dependence of
the eigenvalues of the $3lj3l'j' [J_{1}]2l''j''(J)$  odd-parity
states with $J$=1/2 is given in Figs.~\ref{f-et}a, b, c, and d, where
the energies are
divided by $(Z-3)^2$. Scaled energies $E/(Z-3)^2$
decrease with $Z$ for the first group of levels
(Figs.~\ref{f-et}a, b) and have parabolic $Z$ dependence for the second
group (Figs.~\ref{f-et}c, d). It is evident from
Figs.~\ref{f-et}a, b that there is a strong mixing between
$3s_{1/2}3s_{1/2}[0]2p_{1/2}$, $3s_{1/2}3d_j[J_{1}]2p_{j'}$, and
$3p_{j}3p_{j'}[J_1]2p_{j''}$ levels. We already mentioned that
strong mixing happens between  $3s_{1/2}3s_{1/2}[0]2p_{3/2}$,
$3p_{1/2}3p_{3/2}[1]2p_{3/2}$ and $3p_{1/2}3p_{3/2}[2]2p_{3/2}$
states for $Z$=48--49.

It should be noted that the $LS$ designations are chosen by
comparing results with available experimental data for low-$Z$
ions. We used the labeling of levels from Refs.~\cite{bruch,safr},
where the identification of levels in Fe$^{15+}$ and Cu$^{18+}$
was based on the Cowan code. We already
mentioned that there are no measurements for the second group of
levels, therefore, we do not give $LS$ designations for the levels
shown in Figs.~\ref{f-et}c, d.
The change of relative magnitude of mixing
coefficients is illustrated by the behavior of the six levels
in Fig.~\ref{f-et}a and the five levels in Fig.~\ref{f-et}b.
The crossing of  levels inside of a
complex (set of levels with the same $J$ and parity)
is forbidden by Wigner-Neumann theorem \cite{Landau}.

\section{Comparison of results with other theory and experiment}

We have calculated energies of the 121 odd-parity states
 and the 116 even-parity excited states
for  Na-like ions with
nuclear charges ranging from $Z$=14 to 100.
\begin{table*}
\caption{\label{tab-et}
 Energies of Na-like ions  relative to the ground state of Ne-like ions in
cm$^{-1}$ for ions with $Z$=26-57.}
\begin{ruledtabular}
\begin{tabular}{lrrrrrrrrl}
\multicolumn{1}{c}{$LS$ scheme}&
\multicolumn{1}{c}{$Z$=26}&
\multicolumn{1}{c}{$Z$=29}&
\multicolumn{1}{c}{$Z$=32}&
\multicolumn{1}{c}{$Z$=36}&
\multicolumn{1}{c}{$Z$=42}&
\multicolumn{1}{c}{$Z$=47}&
\multicolumn{1}{c}{$Z$=54}&
\multicolumn{1}{c}{$Z$=57}&
\multicolumn{1}{c}{$jj$ scheme\rule{0ex}{2.3ex}}\\
\hline
         $3s\ ^2S_{1/2}$&-3946654 & -5408870& -7101306& -9720897& -14444875& -19131372& -26886482& -30654172 &$3s_{1/2}$                   \\
$3s3s(^1S)2p\ ^2P_{3/2}$&  1808945&  2258991&  2745096&  3441058&   4560251&   5526519&   6849732&   7377220 &$3s_{1/2}3s_{1/2}[0]2p_{3/2}$\\
$3s3s(^1S)2p\ ^2P_{1/2}$&  1909949&  2425440&  3005011&  3882164&   5436825&   6970009&   9000854&   9844722 &$3s_{1/2}3s_{1/2}[0]2p_{1/2}$\\
$3s3p(^3P)2p\ ^4S_{3/2}$&  2008738&  2500357&  3029307&  3783964&   4993448&   6037431&   7475330&   8054553 &$3s_{1/2}3p_{1/2}[0]2p_{3/2}$\\
$3s3p(^3P)2p\ ^4D_{5/2}$&  2033074&  2527745&  3060001&  3820060&   5040067&   6094393&   7547744&   8133682 &$3s_{1/2}3p_{1/2}[1]2p_{3/2}$\\
$3s3p(^3P)2p\ ^4D_{7/2}$&  2039466&  2544056&  3092521&  3887772&   5202696&   6391332&   8153954&   8929008 &$3s_{1/2}3p_{3/2}[2]2p_{3/2}$\\
$3s3p(^3P)2p\ ^4D_{3/2}$&  2039528&  2535211&  3068422&  3830075&   5053095&   6110756&   7569587&   8158101 &$3s_{1/2}3p_{1/2}[1]2p_{3/2}$\\
$3s3p(^1P)2p\ ^2P_{1/2}$&  2050246&  2548295&  3083346&  3846505&   5069726&   6126870&   7585055&   8173589 &$3s_{1/2}3p_{1/2}[1]2p_{3/2}$\\
$3s3p(^1P)2p\ ^2D_{3/2}$&  2064245&  2571269&  3121744&  3919367&   5237983&   6430207&   8198752&   8976653 &$3s_{1/2}3p_{3/2}[1]2p_{3/2}$\\
$3s3p(^3P)2p\ ^4P_{5/2}$&  2064483&  2573798&  3127096&  3928976&   5254252&   6451971&   8228075&   9009218 &$3s_{1/2}3p_{3/2}[1]2p_{3/2}$\\
$3s3p(^1P)2p\ ^2S_{1/2}$&  2081011&  2593540&  3150111&  3956604&   5289060&   6493231&   8278916&   9064324 &$3s_{1/2}3p_{3/2}[1]2p_{3/2}$\\
$3s3p(^3P)2p\ ^4D_{1/2}$&  2129621&  2676539&  3264927&  4098156&   5504613&   6756597&   8605934&   9416474 &$3s_{1/2}3p_{3/2}[2]2p_{3/2}$\\
$3s3p(^3P)2p\ ^4P_{1/2}$&  2135848&  2685223&  3305872&  4239177&   5881367&   7491636&   9880164&  10785869 &$3s_{1/2}3p_{1/2}[0]2p_{1/2}$\\
$3s3p(^3P)2p\ ^4D_{3/2}$&  2140088&  2685129&  3252230&  4070053&   5417213&   6632701&   8434674&   9227490 &$3s_{1/2}3p_{3/2}[2]2p_{3/2}$\\
$3s3p(^3P)2p\ ^2D_{5/2}$&  2148356&  2668661&  3231730&  4045552&   5387153&   6598225&   8393917&   9183909 &$3s_{1/2}3p_{3/2}[2]2p_{3/2}$\\
$3s3p(^3P)2p\ ^4P_{3/2}$&  2151490&  2701237&  3326210&  4266154&   5921839&   7543576&   9927031&  10836191 &$3s_{1/2}3p_{1/2}[1]2p_{1/2}$\\
$3s3p(^1P)2p\ ^2D_{5/2}$&  2160829&  2733488&  3378432&  4358921&   6115991&   7878203&   9971617&  10881891 &$3s_{1/2}3p_{3/2}[2]2p_{1/2}$\\
$3s3p(^1P)2p\ ^2P_{3/2}$&  2165759&  2724616&  3366873&  4344602&   6098245&   7861912&  10139433&  11066015 &$3s_{1/2}3p_{3/2}[1]2p_{1/2}$\\
$3s3p(^3P)2p\ ^2P_{1/2}$&  2234248&  2803037&  3437436&  4387010&   6072516&   7718801&  10074749&  10996341 &$3s_{1/2}3p_{1/2}[1]2p_{1/2}$\\
$3s3p(^3P)2p\ ^2D_{3/2}$&  2253626&  2839597&  3496708&  4490270&   6270868&   7858384&  10205550&  11168436 &$3s_{1/2}3p_{3/2}[2]2p_{1/2}$\\
$3s3p(^3P)2p\ ^2S_{1/2}$&  2299453&  2882506&  3533957&  4517384&   6280342&   7824403&  10160693&  11394844 &$3s_{1/2}3p_{3/2}[1]2p_{1/2}$\\
$3p3p(^3P)2p\ ^4P_{3/2}$&  2323992&  2875690&  3466402&  4305185&   5644107&   6799972&   8402456&   9055318 &$3p_{1/2}3p_{1/2}[0]2p_{3/2}$\\
$3s3d(^1D)2p\ ^2P_{1/2}$&  2326176&  2884880&  3488911&  4360343&   5793880&   7086329&   9174130&  10022351 &$3s_{1/2}3d_{3/2}[1]2p_{3/2}$\\
$3p3p(^3P)2p\ ^4P_{5/2}$&  2331859&  2890891&  3494970&  4366156&   5799124&   7090560&   9005390&   9849121 &$3p_{1/2}3p_{3/2}[1]2p_{3/2}$\\
$3p3p(^1D)2p\ ^2F_{7/2}$&  2340430&  2900161&  3504745&  4375109&   5807350&   7097923&   9010959&   9853573 &$3p_{1/2}3p_{3/2}[2]2p_{3/2}$\\
$3p3p(^3P)2p\ ^4P_{3/2}$&  2343264&  2903098&  3506655&  4376509&   5807902&   7098885&   9013658&   9857413 &$3p_{1/2}3p_{3/2}[1]2p_{3/2}$\\
$3p3p(^3P)2p\ ^4P_{1/2}$&  2360263&  2934357&  3558208&  4462832&   5937498&   7244558&   9239436&  10097513 &$3p_{1/2}3p_{3/2}[1]2p_{3/2}$\\
$3p3p(^3P)2p\ ^2D_{3/2}$&  2360688&  2923345&  3530947&  4407111&   5848142&   7146773&   9072189&   9920622 &$3p_{1/2}3p_{3/2}[2]2p_{3/2}$\\
$3p3p(^1D)2p\ ^2D_{5/2}$&  2361387&  2925210&  3533627&  4410325&   5851442&   7149759&   9074248&   9921995 &$3p_{1/2}3p_{3/2}[2]2p_{3/2}$\\
$3p3p(^3P)2p\ ^4D_{7/2}$&  2366738&  2941047&  3566519&  4478642&   5996208&   7322997&   9275443&  10133874 &$3p_{3/2}3p_{3/2}[2]2p_{3/2}$\\
$3p3p(^3P)2p\ ^4D_{5/2}$&  2367125&  2941219&  3566447&  4478544&   6004427&   7321838&   9262357&  10114485 &$3p_{3/2}3p_{3/2}[2]2p_{3/2}$\\
$3p3p(^3P)2p\ ^4D_{1/2}$&  2412003&  2993290&  3608811&  4494602&   5959608&   7285858&   9436271&  10311330 &$3p_{1/2}3p_{3/2}[2]2p_{3/2}$\\
$3p3p(^3P)2p\ ^4S_{3/2}$&  2419891&  3004313&  3622761&  4511681&   5970412&   7282549&   9224215&  10078225 &$3p_{3/2}3p_{3/2}[2]2p_{3/2}$\\
$3s3d(^3D)2p\ ^4P_{1/2}$&  2423029&  2998759&  3630191&  4541409&   6064401&   7447505&   9514794&  10579279 &$3p_{3/2}3p_{3/2}[2]2p_{3/2}$\\
$3s3d(^3D)2p\ ^4P_{3/2}$&  2432332&  3009247&  3646026&  4568623&   6048355&   7371746&   9328624&  10189264 &$3p_{3/2}3p_{3/2}[0]2p_{3/2}$\\
$3s3d(^3D)2p\ ^4F_{9/2}$&  2441515&  3018138&  3640285&  4537586&   6017419&   7358627&   9367047&  10262012 &$3s_{1/2}3d_{5/2}[3]2p_{3/2}$\\
$3p3p(^3P)2p\ ^4D_{3/2}$&  2442028&  3044141&  3676510&  4576282&   6090589&   7459516&   9498136&  10404492 &$3s_{1/2}3d_{3/2}[1]2p_{3/2}$\\
$3s3d(^3D)2p\ ^4P_{5/2}$&  2444121&  3022154&  3645035&  4540867&   6009260&   7354901&   9320567&  10184749 &$3s_{1/2}3d_{3/2}[1]2p_{3/2}$\\
$3p3p(^1D)2p\ ^2F_{5/2}$&  2446690&  3035904&  3658924&  4555153&   6029241&   7401210&   9431150&  10330672 &$3s_{1/2}3d_{3/2}[2]2p_{3/2}$\\
$3s3d(^3D)2p\ ^4F_{7/2}$&  2448304&  3025385&  3646951&  4541206&   6018858&   7410665&   9465885&  10368561 &$3s_{1/2}3d_{3/2}[2]2p_{3/2}$\\
$3p3p(^3P)2p\ ^2S_{1/2}$&  2450177&  3066099&  3702033&  4615334&   6110247&   7495342&   9669557&  10796904 &$3p_{1/2}3p_{1/2}[0]2p_{1/2}$\\
$3s3d(^3D)2p\ ^4F_{5/2}$&  2457171&  3059907&  3688115&  4592423&   6082264&   7445653&   9536164&  10459319 &$3s_{1/2}3d_{5/2}[2]2p_{3/2}$\\
$3p3p(^1S)2p\ ^2P_{3/2}$&  2458536&  3050488&  3688938&  4624648&   6178953&   7590274&   9666579&  10585200 &$3s_{1/2}3d_{3/2}[2]2p_{3/2}$\\
$3s3d(^3D)2p\ ^4F_{3/2}$&  2467888&  3072651&  3726831&  4658307&   6214029&   7658925&   9924797&  10978300 &$3s_{1/2}3d_{5/2}[2]2p_{3/2}$\\
$3s3d(^3D)2p\ ^4D_{7/2}$&  2473728&  3057201&  3686257&  4592770&   6086770&   7440067&   9484093&  10396064 &$3s_{1/2}3d_{5/2}[2]2p_{3/2}$\\
$3p3p(^1D)2p\ ^2D_{3/2}$&  2474319&  3087038&  3769447&  4742709&   6305020&   7764437&  10055141&  11119321 &$3s_{1/2}3d_{5/2}[3]2p_{3/2}$\\
$3s3d(^1D)2p\ ^2P_{1/2}$&  2475520&  3073042&  3769007&  4724883&   6270683&   7669065&   9823175&  10854112 &$3s_{1/2}3d_{3/2}[2]2p_{3/2}$\\
$3s3d(^1D)2p\ ^2F_{5/2}$&  2476256&  3072416&  3770426&  4716884&   6243639&   7642346&   9827384&  10855818 &$3s_{1/2}3d_{5/2}[3]2p_{3/2}$\\
$3p3p(^3P)2p\ ^2D_{5/2}$&  2478055&  3117666&  3792216&  4823275&   6692492&   8552527&  11702449&  12906145 &$3p_{1/2}3p_{3/2}[2]2p_{1/2}$\\
$3s3d(^3D)2p\ ^2P_{3/2}$&  2494618&  3110698&  3801760&  4822655&   6690006&   8549142&  11693655&  12876722 &$3p_{1/2}3p_{3/2}[1]2p_{1/2}$\\
$3s3d(^3D)2p\ ^4D_{1/2}$&  2503765&  3111834&  3774944&  4819276&   6600142&   8335220&  11178869&  12561626 &$3s_{1/2}3d_{5/2}[2]2p_{3/2}$\\
$3s3d(^3D)2p\ ^4D_{3/2}$&  2532307&  3146832&  3830007&  4918354&   6841218&   8727106&  11696997&  13061250 &$3p_{1/2}3p_{3/2}[2]2p_{1/2}$\\
$3s3d(^3D)2p\ ^2F_{7/2}$&  2536607&  3128717&  3766505&  4684639&   6205225&   7605894&   9808688&  10842983 &$3s_{1/2}3d_{5/2}[3]2p_{3/2}$\\
$3s3d(^3D)2p\ ^4D_{5/2}$&  2553254&  3150273&  3837776&  4932054&   6886124&   8784045&  11721257&  13062930 &$3p_{3/2}3p_{3/2}[2]2p_{1/2}$\\
$3s3d(^3D)2p\ ^4F_{3/2}$&  2553625&  3193957&  3905127&  4973425&   6875056&   8755816&  11862792&  13125630 &$3p_{3/2}3p_{3/2}[2]2p_{1/2}$\\
$3s3d(^1D)2p\ ^2D_{5/2}$&  2554863&  3199650&  3916483&  4994914&   6912385&   8834753&  11836413&  13135682 &$3s_{1/2}3d_{3/2}[2]2p_{1/2}$\\
$3p3p(^1S)2p\ ^2P_{1/2}$&  2560337&  3171081&  3838660&  4841463&   6685037&   8542196&  11682890&  12867258 &$3p_{1/2}3p_{3/2}[1]2p_{1/2}$\\
\end{tabular}
\end{ruledtabular}
\end{table*}
In Table~\ref{tab-et},
we illustrate our theoretical results giving energies of the 38
odd-parity states
$3s^2(^1S)2p\ ^2P_{J}$, $3s3d(^{1,3}D)2p\ ^{2,4}L_{J}$, and
$3p^2(^{1,3}L')2p\ ^{2,4}L_{J}$  and the
18 even-parity states
$3s3p(^{1,3}P)2p\ ^{2,4}L_{J}$ for
eight Na-like ions
with nuclear charges ranging from $Z$=26 to $Z$=57.
We also include the ground state $3s\ ^2S_{1/2}$ energies
in the table.

\subsection{Excitation energies}

Comparisons of our RMBPT energies with other theoretical and
experimental data are too voluminous to include here; therefore,
we present several examples of comparisons for selected levels and ions.
\begin{table}
\caption{\label{tab-com1}
 Auger energies  in Fe$^{15+}$ in eV.
Comparison with predicted data ($E_{\rm fit}$) from  Bliman {\it
et al.\/} \protect\cite{a5}.}
\begin{ruledtabular}
\begin{tabular}{llll}
\multicolumn{1}{c}{$LS$ scheme}& \multicolumn{1}{c} {$E_{\rm
RMBPT}$} & \multicolumn{1}{c} {$E_{\rm fit}$} & \multicolumn{1}{c}{$jj$
scheme
\rule{0ex}{2.3ex}}\\
\hline
$3s3s(^1S)2p\ ^2P_{3/2}$&   224.3&  225.0&   $3s_{1/2}3s_{1/2}[0]2p_{3/2}$\\
$3s3s(^1S)2p\ ^2P_{1/2}$&   236.8&  237.3&   $3s_{1/2}3s_{1/2}[0]2p_{1/2}$\\
$3s3p(^3P)2p\ ^4S_{3/2}$&   249.1&  249.5&   $3s_{1/2}3p_{1/2}[0]2p_{3/2}$\\
$3s3p(^3P)2p\ ^4D_{5/2}$&   252.1&  252.4&   $3s_{1/2}3p_{1/2}[1]2p_{3/2}$\\
$3s3p(^3P)2p\ ^4D_{7/2}$&   252.9&  253.2&   $3s_{1/2}3p_{3/2}[2]2p_{3/2}$\\
$3s3p(^3P)2p\ ^4D_{3/2}$&   252.9&  253.2&   $3s_{1/2}3p_{1/2}[1]2p_{3/2}$\\
$3s3p(^1P)2p\ ^2P_{1/2}$&   254.2&  254.6&   $3s_{1/2}3p_{1/2}[1]2p_{3/2}$\\
$3s3p(^1P)2p\ ^2D_{3/2}$&   255.9&  256.0&   $3s_{1/2}3p_{3/2}[1]2p_{3/2}$\\
$3s3p(^3P)2p\ ^4P_{5/2}$&   256.0&  256.1&   $3s_{1/2}3p_{3/2}[1]2p_{3/2}$\\
$3s3p(^1P)2p\ ^2S_{1/2}$&   258.0&  258.3&   $3s_{1/2}3p_{3/2}[1]2p_{3/2}$\\
$3s3p(^3P)2p\ ^4D_{1/2}$&   264.0&  264.5&   $3s_{1/2}3p_{3/2}[2]2p_{3/2}$\\
$3s3p(^3P)2p\ ^4P_{1/2}$&   264.8&  265.1&   $3s_{1/2}3p_{1/2}[0]2p_{1/2}$\\
$3s3p(^3P)2p\ ^4D_{3/2}$&   265.3&  265.7&   $3s_{1/2}3p_{3/2}[2]2p_{3/2}$\\
$3s3p(^3P)2p\ ^2D_{5/2}$&   266.4&  266.5&   $3s_{1/2}3p_{3/2}[2]2p_{3/2}$\\
$3s3p(^3P)2p\ ^4P_{3/2}$&   266.8&  267.1&   $3s_{1/2}3p_{1/2}[1]2p_{1/2}$\\
$3s3p(^1P)2p\ ^2D_{5/2}$&   267.9&  268.1&   $3s_{1/2}3p_{3/2}[2]2p_{1/2}$\\
$3s3p(^1P)2p\ ^2P_{3/2}$&   268.5&  268.5&   $3s_{1/2}3p_{3/2}[1]2p_{1/2}$\\
$3s3p(^3P)2p\ ^2P_{1/2}$&   277.0&  277.1&   $3s_{1/2}3p_{1/2}[1]2p_{1/2}$\\
$3s3p(^3P)2p\ ^2D_{3/2}$&   279.4&  279.6&   $3s_{1/2}3p_{3/2}[2]2p_{1/2}$\\
$3s3p(^3P)2p\ ^2S_{1/2}$&   285.1&  285.1&   $3s_{1/2}3p_{3/2}[1]2p_{1/2}$\\
$3p3p(^3P)2p\ ^4P_{3/2}$&   288.1&  289.2&   $3p_{1/2}3p_{1/2}[0]2p_{3/2}$\\
$3s3d(^1D)2p\ ^2P_{1/2}$&   288.4&  289.6&   $3s_{1/2}3d_{3/2}[1]2p_{3/2}$\\
$3p3p(^3P)2p\ ^4P_{5/2}$&   289.1&  290.0&   $3p_{1/2}3p_{3/2}[1]2p_{3/2}$\\
$3p3p(^1D)2p\ ^2F_{7/2}$&   290.2&  291.3&   $3p_{1/2}3p_{3/2}[2]2p_{3/2}$\\
$3p3p(^3P)2p\ ^4P_{3/2}$&   290.5&       &   $3p_{1/2}3p_{3/2}[1]2p_{3/2}$\\
$3p3p(^3P)2p\ ^4P_{1/2}$&   292.6&       &   $3p_{1/2}3p_{3/2}[1]2p_{3/2}$\\
$3p3p(^3P)2p\ ^2D_{3/2}$&   292.7&  293.3&   $3p_{1/2}3p_{3/2}[2]2p_{3/2}$\\
$3p3p(^1D)2p\ ^2D_{5/2}$&   292.8&       &   $3p_{1/2}3p_{3/2}[2]2p_{3/2}$\\
$3p3p(^3P)2p\ ^4D_{7/2}$&   293.4&  294.1&   $3p_{3/2}3p_{3/2}[2]2p_{3/2}$\\
$3p3p(^3P)2p\ ^4D_{5/2}$&   293.5&  293.6&   $3p_{3/2}3p_{3/2}[2]2p_{3/2}$\\
$3p3p(^3P)2p\ ^4D_{1/2}$&   299.1&       &   $3p_{1/2}3p_{3/2}[2]2p_{3/2}$\\
$3p3p(^3P)2p\ ^4S_{3/2}$&   300.0&  300.4&   $3p_{3/2}3p_{3/2}[2]2p_{3/2}$\\
$3s3d(^3D)2p\ ^4P_{1/2}$&   300.4&       &   $3p_{3/2}3p_{3/2}[2]2p_{3/2}$\\
$3s3d(^3D)2p\ ^4P_{3/2}$&   301.6&       &   $3p_{3/2}3p_{3/2}[0]2p_{3/2}$\\
$3s3d(^3D)2p\ ^4F_{9/2}$&   302.7&  303.6&   $3s_{1/2}3d_{5/2}[3]2p_{3/2}$\\
$3p3p(^3P)2p\ ^4D_{3/2}$&   302.8&       &   $3s_{1/2}3d_{3/2}[1]2p_{3/2}$\\
$3s3d(^3D)2p\ ^4P_{5/2}$&   303.0&  303.7&   $3s_{1/2}3d_{3/2}[1]2p_{3/2}$\\
$3p3p(^1D)2p\ ^2F_{5/2}$&   303.4&       &   $3s_{1/2}3d_{3/2}[2]2p_{3/2}$\\
$3s3d(^3D)2p\ ^4F_{7/2}$&   303.6&  304.4&   $3s_{1/2}3d_{3/2}[2]2p_{3/2}$\\
$3p3p(^3P)2p\ ^2S_{1/2}$&   303.8&       &   $3p_{1/2}3p_{1/2}[0]2p_{1/2}$\\
$3s3d(^3D)2p\ ^4F_{5/2}$&   304.7&  305.3&   $3s_{1/2}3d_{5/2}[2]2p_{3/2}$\\
$3p3p(^1S)2p\ ^2P_{3/2}$&   304.8&       &   $3s_{1/2}3d_{3/2}[2]2p_{3/2}$\\
$3s3d(^3D)2p\ ^4F_{3/2}$&   306.0&       &   $3s_{1/2}3d_{5/2}[2]2p_{3/2}$\\
$3s3d(^3D)2p\ ^4D_{7/2}$&   306.7&       &   $3s_{1/2}3d_{5/2}[2]2p_{3/2}$\\
$3p3p(^1D)2p\ ^2D_{3/2}$&   306.8&       &   $3s_{1/2}3d_{5/2}[3]2p_{3/2}$\\
$3s3d(^1D)2p\ ^2P_{1/2}$&   306.9&       &   $3s_{1/2}3d_{3/2}[2]2p_{3/2}$\\
$3s3d(^1D)2p\ ^2F_{5/2}$&   307.0&       &   $3s_{1/2}3d_{5/2}[3]2p_{3/2}$\\
$3p3p(^3P)2p\ ^2D_{5/2}$&   307.2&       &   $3p_{1/2}3p_{3/2}[2]2p_{1/2}$\\
$3s3d(^3D)2p\ ^2P_{3/2}$&   309.3&       &   $3p_{1/2}3p_{3/2}[1]2p_{1/2}$\\
$3s3d(^3D)2p\ ^4D_{1/2}$&   310.4&       &   $3s_{1/2}3d_{5/2}[2]2p_{3/2}$\\
$3s3d(^3D)2p\ ^4D_{3/2}$&   314.0&       &   $3p_{1/2}3p_{3/2}[2]2p_{1/2}$\\
$3s3d(^3D)2p\ ^2F_{7/2}$&   314.5&       &   $3s_{1/2}3d_{5/2}[3]2p_{3/2}$\\
$3s3d(^3D)2p\ ^4D_{5/2}$&   316.6&       &   $3p_{3/2}3p_{3/2}[2]2p_{1/2}$\\
$3s3d(^3D)2p\ ^4F_{3/2}$&   316.6&  317.1&   $3p_{3/2}3p_{3/2}[2]2p_{1/2}$\\
$3s3d(^1D)2p\ ^2D_{5/2}$&   316.8&       &   $3s_{1/2}3d_{3/2}[2]2p_{1/2}$\\
$3p3p(^1S)2p\ ^2P_{1/2}$&   317.4&       &   $3p_{1/2}3p_{3/2}[1]2p_{1/2}$\\
$3p3p(^3P)2p\ ^2P_{1/2}$&   317.5&       &   $3p_{3/2}3p_{3/2}[0]2p_{1/2}$\\
\end{tabular}
\end{ruledtabular}
\end{table}
 In Table~\ref{tab-com1},
 our theoretical results for
$3s^2(^1S)2p\ ^2P_{J}$, $3s3p(^{1,3}P)2p\ ^{2,4}L_{J}$,
$3p^2(^{1,3}L')2p\ ^{2,4}L_{J}$, and $3s3d(^{1,3}D)2p\
^{2,4}L_{J}$ levels  of Na-like Fe are compared with predicted
Auger energies given by \citet{a5}.
It should be noted that the predicted Auger energies in
Ref.~\cite{a5} differ by 1~eV from experimental measurements
presented more than 15 years ago by \citet{a3}, where the method of zero-degree Auger
spectroscopy was used to study the excitation and decay of Auger
states formed in 170-keV Fe$^{17+}$ ions with He and Ne target
atoms. Our RMBPT energies are in much better agreement with recent
results from \cite{a5} than with experimental measurements
from \cite{a3}.

\begin{table}
\caption{\label{tab-com2}
 Wavelengths $\lambda$ in \AA$\,$ for Na-like ions for odd-parity states
 given relative to the ground state. Comparison with experimental results from
  Ref.~\protect\cite{b86} ($\lambda_a$) and Ref.~\protect\cite{b95} ($\lambda_b$).}
\begin{ruledtabular}
\begin{tabular}{llllll}
\multicolumn{2}{c} {}
& \multicolumn{2}{c} {$Z$=47}&
\multicolumn{2}{c} {$Z$=57}\\
\multicolumn{1}{c}{Configuration}& \multicolumn{1}{c}{J}&
\multicolumn{1}{c} {$\lambda_{\rm RMBPT}$} & \multicolumn{1}{c}
{$\lambda_{a}$} & \multicolumn{1}{c} {$\lambda_{\rm RMBPT}$} &
\multicolumn{1}{c} {$\lambda_{b}$} \\\\
\hline
$3s_{1/2}3s_{1/2}[0]2p_{3/2}$& 1.5&     4.0555& 4.0547&  2.6294&  2.6294\\
$3s_{1/2}3s_{1/2}[0]2p_{1/2}$& 0.5&     3.8312& 3.8309&  2.4692&    \\
$3s_{1/2}3d_{3/2}[2]2p_{3/2}$& 1.5&     3.7423& 3.7427&  2.4249&    \\
$3s_{1/2}3d_{5/2}[2]2p_{1/2}$& 2.5&     3.5707& 3.571 &  2.2788&    \\
$3p_{3/2}3p_{3/2}[0]2p_{1/2}$& 0.5&     3.5701& 3.571 &  2.2798&    \\
$3s_{1/2}3d_{3/2}[2]2p_{1/2}$& 1.5&     3.5700& 3.571 &  2.2766&    \\
$3s_{1/2}3p_{3/2}[1]2s_{1/2}$& 1.5&     3.4261& 3.426 &  2.1805&    \\
$3s_{1/2}3p_{3/2}[1]2s_{1/2}$& 0.5&     3.4121& 3.412 &  2.1739&    \\
\end{tabular}
\end{ruledtabular}
\end{table}
In Table~\ref{tab-com2}, our RMBPT wavelengths
  for Na-like Ag and La are compared
  with experimental results presented by \citet{b86,b95}.
  Spectra of the $n$=3 to $n$=2
 transitions in neonlike silver obtained from the Princeton Large
 Torus were recorded with high-resolution Bragg-crystal
 spectrometer. The measurements covered the wavelength interval
 3.3--4.1~\AA. Eight lines of Na-like Ag
 were observed.
 As one can see from Table~\ref{tab-com2}, our theoretical
 calculations are in excellent agreement with results from
 Ref.~\cite{b86}.
  Measurement of a single line in  Na-like La was  carried out with
high-resolution crystal spectrometer on the PLT tokamak by \citet{b95}.
Our RMBPT result agrees with experimental wavelengths from \cite{b95} to
five significant figures. We list our theoretical data for
other  lines in La$^{46+}$ in Table~\ref{tab-com2}.

\begin{table}
\caption{\label{tab-com3}
 Wavelengths $\lambda$ in \AA$\,$  for the $3s3p(^3P)2p\
^4D_{7/2}$--$3s3d(^3D)2p\ ^4F_{9/2}$  transition
 in Na-like ions
 with $Z$=15-30. Comparison with experimental ($\lambda_{\rm expt}$)
and predicted ($\lambda_{\rm fit}$) data  from  Jupen {\it et
al.\/} \protect\cite{a2}.}
\begin{ruledtabular}
\begin{tabular}{llll}
\multicolumn{1}{c}{$Z$}& \multicolumn{1}{c} {$\lambda_{\rm RMBPT}$}
& \multicolumn{1}{c} {$\lambda_{\rm expt}$} &
\multicolumn{1}{c}{$\lambda_{\rm fit}$}\\\\
\hline
 15&      729.74&                   &       728.3 \\
 16&      613.99&    614.06$\pm$0.08&       614.20\\
 17&      534.12&    532.70$\pm$0.10&       532.67\\
 18&      472.79&    471.34$\pm$0.05&       471.22\\
 19&      424.23&                   &       424.05\\
 20&      384.80&                   &       384.15\\
 21&      350.86&                   &       351.98\\
 22&      325.76&    324.87$\pm$0.02&       324.89\\
 23&      302.61&                   &       301.69\\
 24&      281.75&    281.67$\pm$0.08&       281.60\\
 25&      264.32&                   &       263.98\\
 26&      248.73&    248.36$\pm$0.05&       248.39\\
 27&      234.76&                   &       234.48\\
 28&      222.26&    222.00$\pm$0.10&       221.98\\
 29&      210.93&    210.70$\pm$0.05&       210.68\\
30&      200.63&                   &       200.39\\
\end{tabular}
\end{ruledtabular}
\end{table}
The $3s3p(^3P)2p\
^4D_{7/2}$--$3s3d(^3D)2p\ ^4F_{9/2}$  transitions
 in P$^{4+}$-Zn$^{19+}$ ions
were investigated  by \citet{a2}. Identification of spectral lines
 originating from the transitions between core-excited configurations
 in Na isoelectronic sequence was made from
 analyses of beam-foil spectra in
\cite{a2}.
 Wavelengths
 $\lambda_{\rm expt}$ from \cite{a2}
are shown in Table~\ref{tab-com3} together with our RMBPT data
{$\lambda_{\rm RMBPT}$}.
We also present the  predicted
data $\lambda_{\rm fit}$ from Ref.~\cite{a2} in the same table.
The  predicted
wavelengths were obtained in \cite{a2} from the fit of the
 difference between experimental data and Cowan code
calculations. Our {\it ab initio}
calculations are in good agreement with those predictions.

\subsection{Fine structure of the $^2L$ and $^4L$ terms
 in core-excited states of Na-like ions}
\begin{figure*}
\centerline{
\includegraphics[scale=0.32]{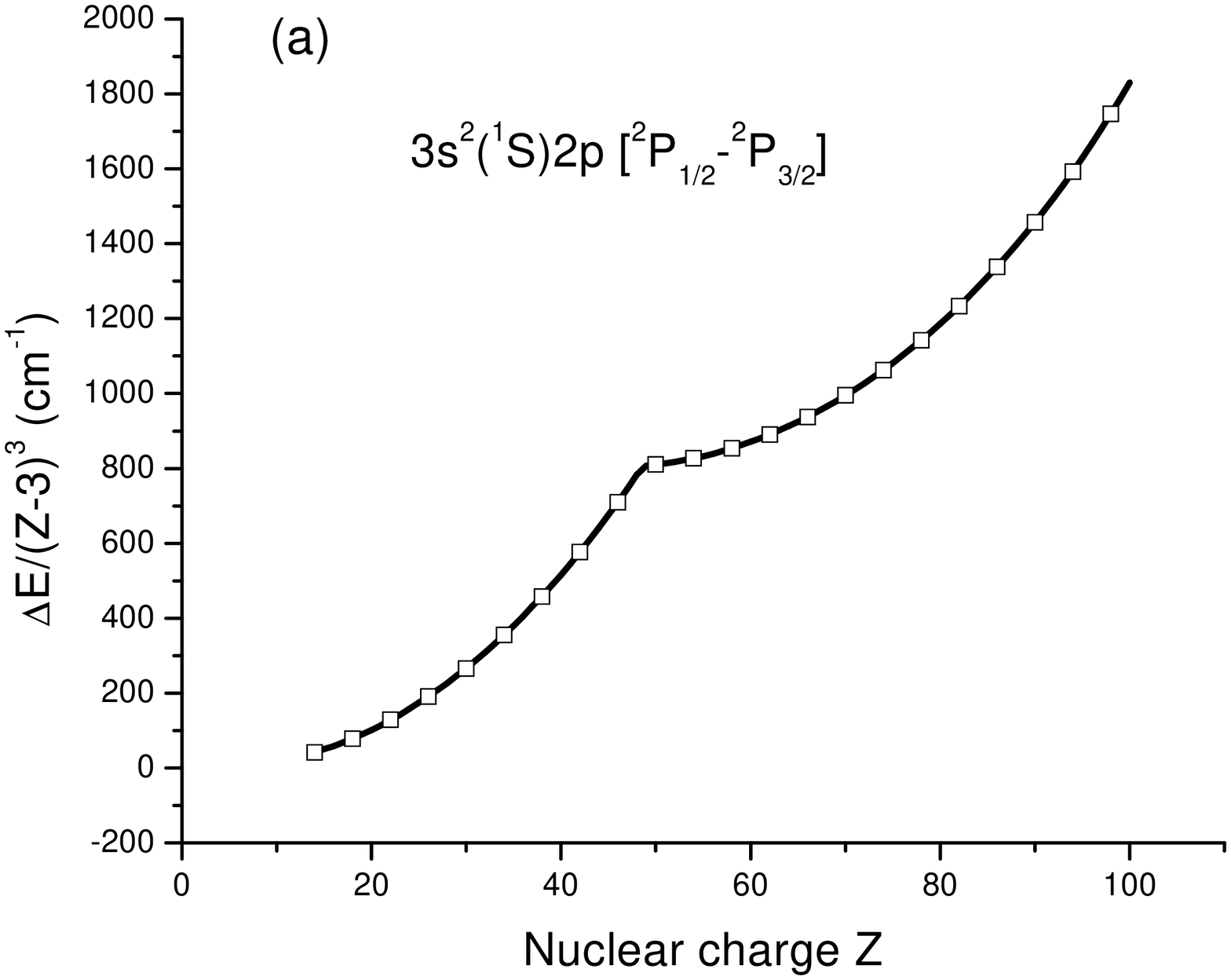}
\includegraphics[scale=0.32]{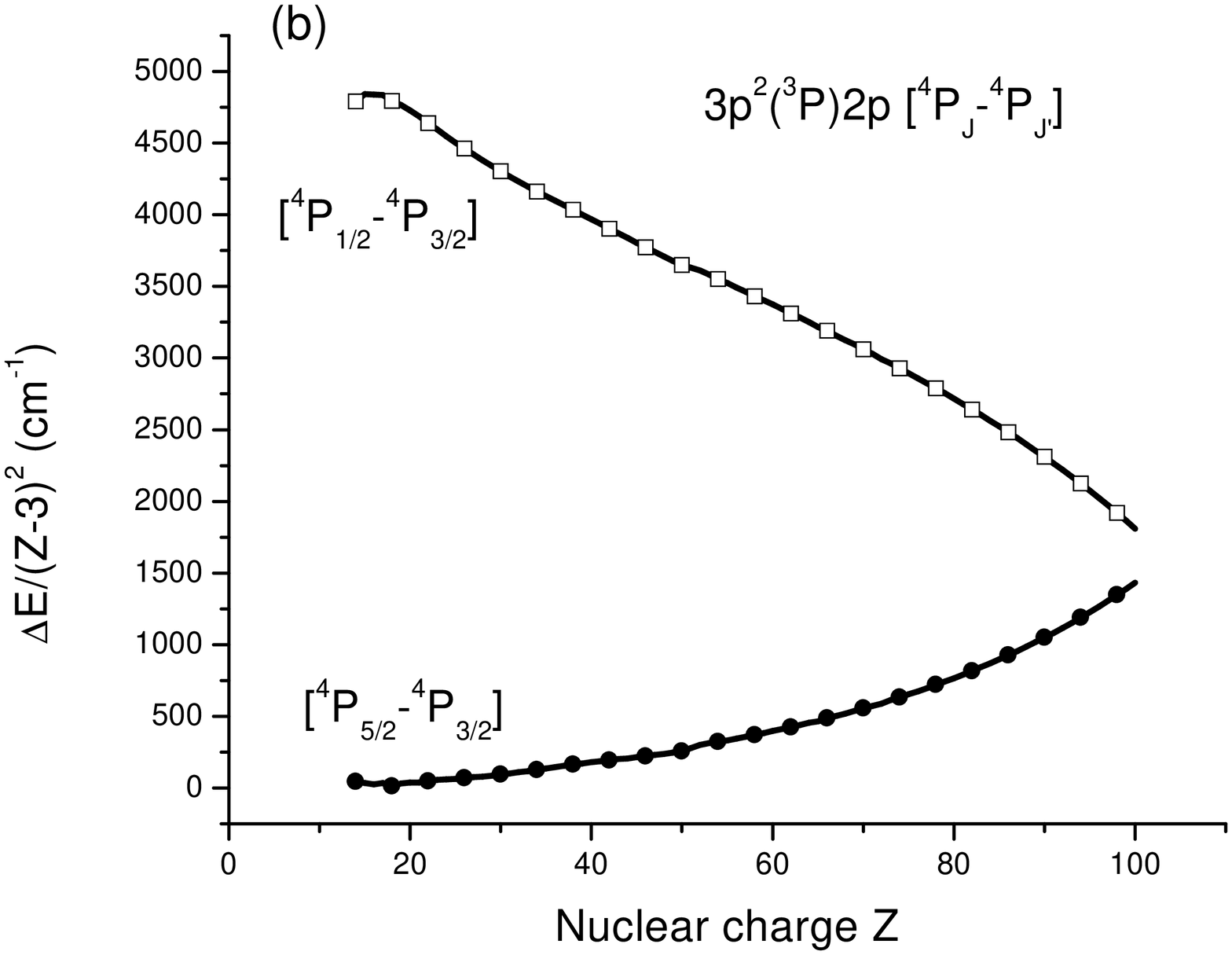}}
\centerline{
\includegraphics[scale=0.32]{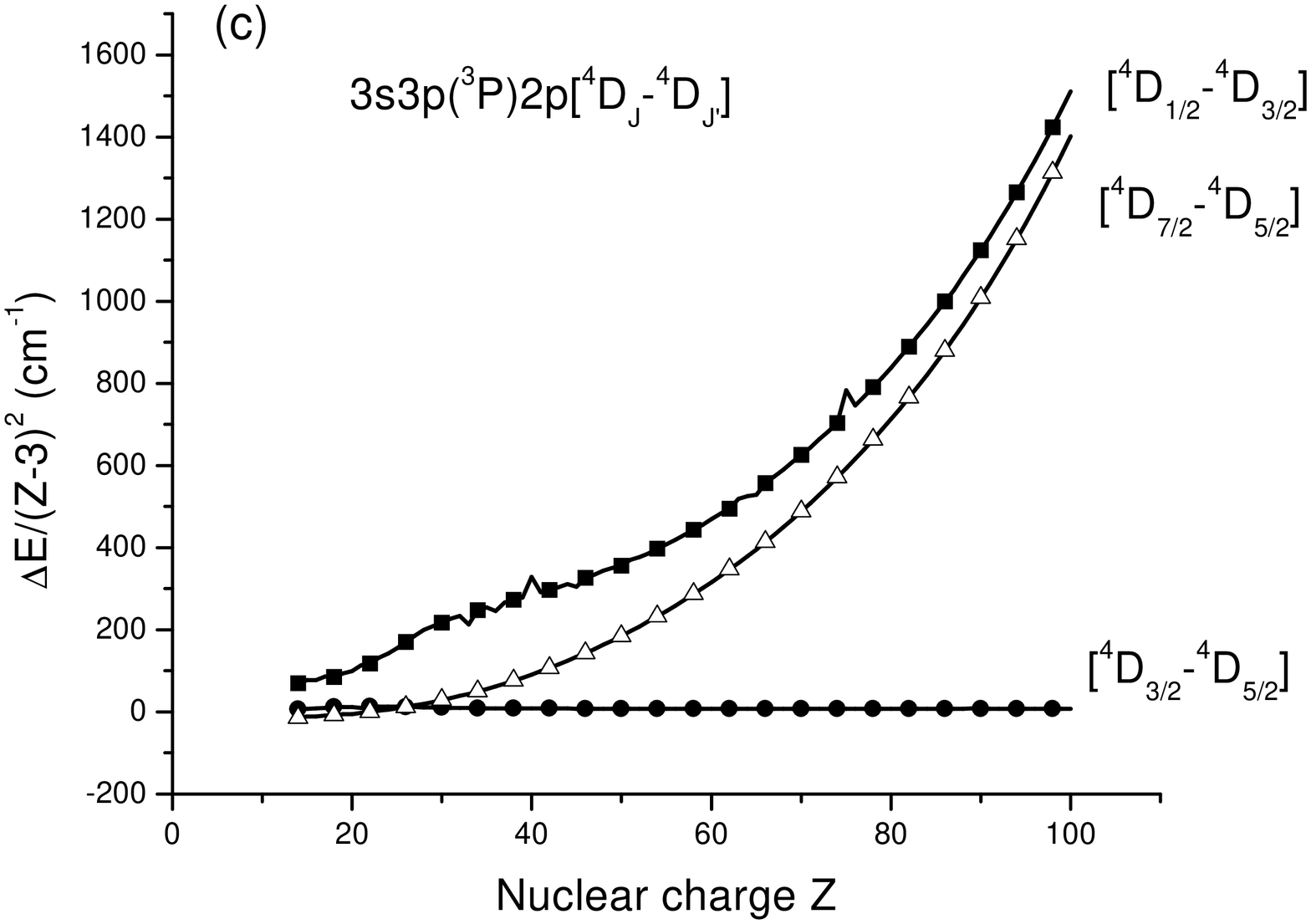}
\includegraphics[scale=0.32]{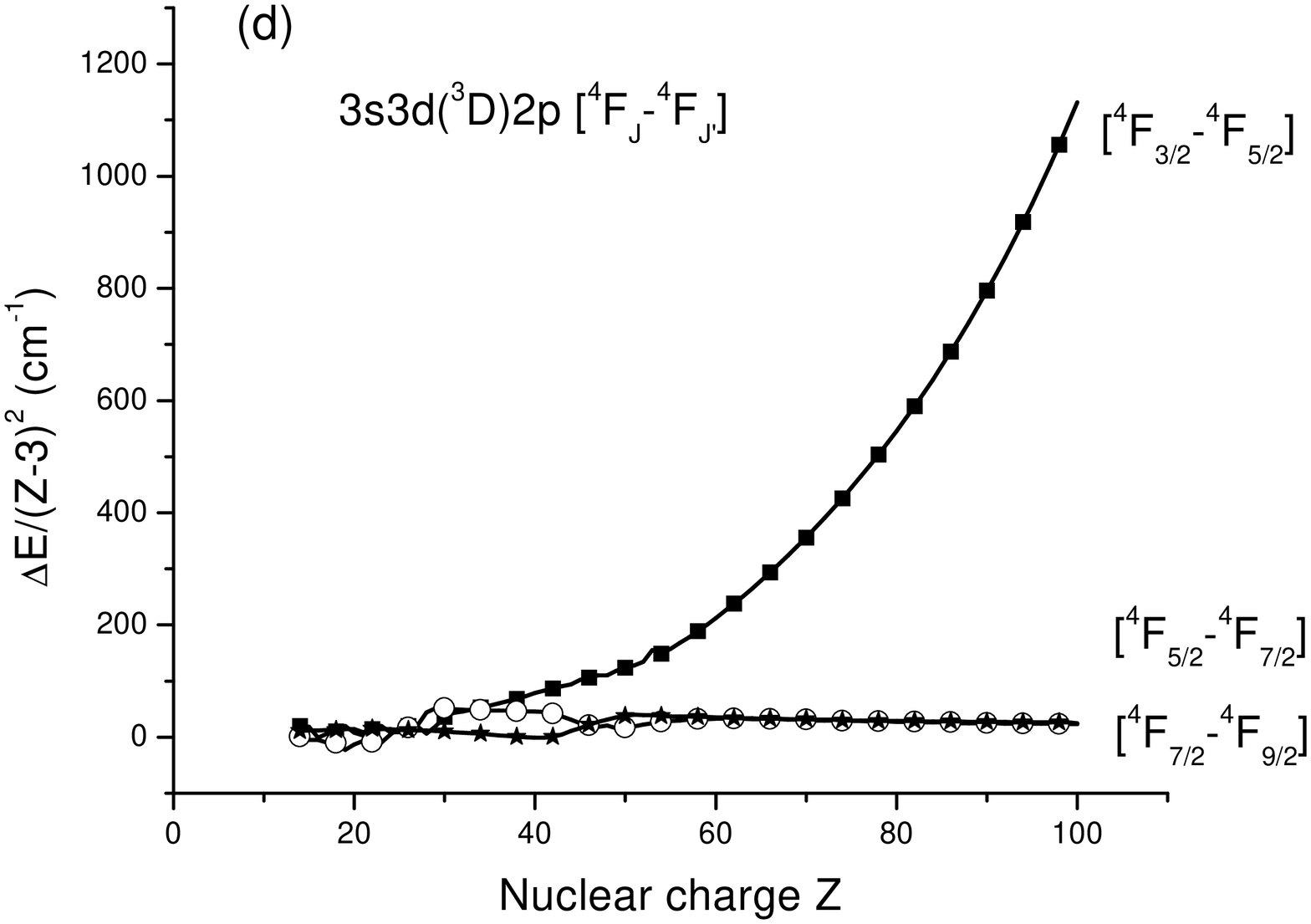}}
\caption{Energy splittings (${\Delta}E/(Z-3)^4$ in cm$^{-1}$) as
functions of $Z$.} \label{split}
\end{figure*}
 No measurements of  fine-structure intervals were
made by observing the wavelength differences between transitions
within the doublet or quartet  states. The intervals are obtained
by subtracting the excitation energies of the corresponding levels
(see, for example, Auger data in Table~\ref{tab-com1}).  These
fine structures are quite regular throughout the isoelectronic
sequence, as seen from Fig.~\ref{split}a. In this figure, we
present the fine-structure splitting divided by $(Z-3)^2$ for the
doublet term $3s^2(^1S)2p\ ^2P$. Small deviation from smooth line
of Fig.~\ref{split}a  in the region of $Z$=48-49 is due to the
strong mixing in the odd-parity complex with $J$=1/2 as
illustrated in Fig.~\ref{contr}a. The $3s^2(^1S)2p\ ^2P_{3/2}$
level has smaller energy than the $3s^2(^1S)2p\ ^2P_{1/2}$ level
in contradiction with Land\'{e} interval rules. The same situation
occurs for other fine-structure intervals  in Figs.~\ref{split}b,
c, and d. The $^4P_{3/2}$ level has the
 smallest energy  among  $3p^2(^3P)2p\
^4P_{J}$  levels, the next level is
$^4P_{5/2}$, and  the $^4P_{1/2}$ level has the largest
energy. The value of $(E_{1/2}-E_{3/2})/(Z-3)^2$  decreases
with $Z$ while the value of $(E_{5/2}-E_{3/2})/(Z-3)^2$  increases
with $Z$. As a result, energies of $^4P_{1/2}$ and
$^4P_{5/2}$ become almost equal for high $Z$.
  The energy difference between the $^4D_{3/2}$ and $^4D_{5/2}$ levels
in the $3s3p(^3P)2p\ ^4D_J$ quartet term is very small, as shown in
 Fig.~\ref{split}c.
 We find that the $^4D_J$ quartet terms are ordered as follows:
 $^4D_{1/2}$, $^4D_{7/2}$, $^4D_{3/2}$, and $^4D_{5/2}$.
Energy differences between the $^4F_{5/2}$, $^4F_{7/2}$, and
$^4F_{9/2}$ levels in the $3s3d(^3D)2p\ ^4F_J$ quartet term are
very small, as shown in
 Fig.~\ref{split}d.
 The unusual splitting is caused by mixing of states
 within the complexes with different $J$.
  Further experimental confirmation would be
very helpful in verifying the correctness of these occasionally
sensitive mixing parameters.

\section{Conclusion}
A systematic second-order RMBPT study of the energies
of the core-excited  states of  Na-like ions has been presented.
RMBPT gives results in good agreement with
experimental and predicted data. It would be beneficial if
experimental data for other highly-charged Na-like ions were
available. At the present time, there are almost no experimental
data between $Z$=47 and $Z$=100 for the sodium isoelectronic
sequence. Availability of such data could lead to an improved
understanding of the relative importance of different
contributions to the energies of highly-charged ions.
 These calculations are presented as a theoretical
benchmark for comparison with experiment and theory. The present
calculations
could be further improved by including third-order correlation
corrections.
\begin{acknowledgments}
The work of W.R.J. and M.S.S. was supported in part by National
Science Foundation Grant No.\ PHY-01-39928. U.I.S. acknowledges
partial support by Grant No.\ B516165 from Lawrence Livermore
National Laboratory. The work of J.R.A was performed under the
auspices of the U. S. Department of Energy by the University of
California, Lawrence Livermore National Laboratory under contract
No.\ W-7405-Eng-48.
\end{acknowledgments}

\begin{widetext}
\appendix*
\section{The particle-particle-hole diagram}
The model space state vector for an ion with two valence electrons
($v, w $)  and one hole ($a$)  can be
represented as \cite{bor2}:
\begin{equation}
\Psi(QJM)=N(Q) \sum \langle v w| K_{12}\rangle \langle K_{12} a| K
\rangle
 a^\dagger_{v} a^\dagger_{w}
a_{a} | 0 \rangle \,, \label{eq1}
\end{equation}
where
 $|0 \rangle$ is the state vector for the core ($1s^22s^22p^6$ in our case),
$Q$ describes a particle-particle-hole state with quantum numbers
$n^0_v\kappa^0_vn^0_w\kappa^0_w n^0_a\kappa^0_a$, and
intermediate momentum $J_{12}$. We use the notation $K_i=\{J_i,
M_i\}$ and $v =\{j_v, m_v\}$. The sum in Eq.(\ref{eq1}) is over
magnetic quantum numbers $m_v$, $m_w$, $m_a$, and $M_{12}$. The
quantity $\langle K_1 K_2|K_3 \rangle$ is a Clebsch-Gordan
coefficient:
\begin{equation}
\langle K_1 K_2| K_3 \rangle=(-1)^{J_1-J_2+M_3} \sqrt{2J_3+1}
\left( \begin{array}{ccc}
    J_1  & J_2  & J_3 \\
    M_1  & M_2  & -M_3
        \end{array} \right) \,.
    \label{CG}
\end{equation}

The above representation of the state vector is somewhat
inconvenient; for example,  it leads to an expression containing
36 terms for the second-order diagram with six free ends. It was shown in
Ref.~\cite{bor2} that  it is more efficient to express the state
vector in a manifestly symmetric form in
the case of three-electron system. Here, we can only use
symmetry of two particles $v$ and $w$. As a result, we obtain the
expression for the particle-particle-hole contribution
$E^{(2)}_3$ in second order consisting of  a sum of $E_a$, $E_b$, $E_c$, and
$E_d$ terms:
$$
E^{(2)}_3=E_a+E_b+E_c+E_d,
$$
\begin{multline}
E_{a}(v^{0}w^{0}[J_{12}]aJ,x^{0}y^{0}[J_{13}]bJ)=-\sum_{vw}\sum_{v^{%
\prime }w^{\prime
}}P_{J_{12}}(v^{0}v,w^{0}w)P_{J_{13}}(x^{0}v^{\prime
},y^{0}w^{\prime }) \\ 
\times
(vw|K_{12})(K_{12}a|K)(v^{\prime }w^{\prime
}|K_{13})(K_{13}b|K)\sum_{n}\frac{g(vwan)g(nbv^{\prime }w^{\prime })}{%
\varepsilon (v)+\varepsilon (w)-\varepsilon (a)-\varepsilon
(n)}\, ,
\end{multline}
\begin{multline}
E_{b}(v^{0}w^{0}[J_{12}]aJ,x^{0}y^{0}[J_{13}]bJ)=\sum_{vw}\sum_{v^{%
\prime }w^{\prime
}}P_{J_{12}}(v^{0}v,w^{0}w)P_{J_{13}}(x^{0}v^{\prime
},y^{0}w^{\prime }) \\
\times
(vw|K_{12})(K_{12}a|K)(v^{\prime }w^{\prime
}|K_{13})(K_{13}b|K)\sum_{n}\left[ \frac{g(vww^{\prime
}n)g(nbv^{\prime }a)-g(vww^{\prime }n)g(nbav^{\prime
})}{\varepsilon (v)+\varepsilon (w)-\varepsilon (w^{\prime
})-\varepsilon (n)}\right]\,,
\end{multline}
\begin{multline}
E_{c}(v^{0}w^{0}[J_{12}]aJ,x^{0}y^{0}[J_{13}]bJ)=\sum_{vw}\sum_{v^{%
\prime }w^{\prime
}}P_{J_{12}}(v^{0}v,w^{0}w)P_{J_{13}}(x^{0}v^{\prime
},y^{0}w^{\prime }) \\
\times
(vw|K_{12})(K_{12}a|K)(v^{\prime }w^{\prime
}|K_{13})(K_{13}b|K)\sum_{n}\left[ \frac{g(vban)g(nwv^{\prime
}w^{\prime })-g(bvan)g(nwv^{\prime }w^{\prime })}{\varepsilon
(v)+\varepsilon (b)-\varepsilon (a)-\varepsilon (n)}\right]\,,
\end{multline}
\begin{multline}
E_{d}(v^{0}w^{0}[J_{12}]aJ,x^{0}y^{0}[J_{13}]bJ)=\sum_{vw}\sum_{v^{%
\prime }w^{\prime
}}P_{J_{12}}(v^{0}v,w^{0}w)P_{J_{13}}(x^{0}v^{\prime
},y^{0}w^{\prime }) \\
 =\sum_{vw}\sum_{v^{\prime
}w^{\prime }}(vw|K_{12})(K_{12}a|K)(v^{\prime }w^{\prime
}|K_{13})(K_{13}b|K)\sum_{n}\frac{1}{\varepsilon
(w)+\varepsilon (b)-\varepsilon (v^{\prime })-\varepsilon (n)}
\\
 \times \left[ -g(wbv^{\prime }n)g(nvw^{\prime
}a)+g(bwv^{\prime }n)g(nvw^{\prime }a)+g(wbv^{\prime
}n)g(nvaw^{\prime })-g(bwv^{\prime }n)g(nvaw^{\prime })\right]\,.
\end{multline}
The quantity $P_{J}(v^{0}v,w^{0}w)$ is a symmetry coefficient
defined by
\begin{equation}
P_{J}(v^{0}v,w^{0}w)=\eta _{v^{0}w^{0}}\left[ \delta
_{v^{0}v}\delta _{w^{0}w}+(-1)^{j_{v}+j_{w}+J+1}\delta
_{v^{0}w}\delta _{w^{0}v}\right] \,,
\end{equation}
where $\eta _{vw}$ is a normalization factor given by
\[
\eta _{vw}=\left\{
\begin{array}{ll}
1 & \mbox{for $w \neq v$} \\
1/\sqrt{2} & \mbox{for $w = v$}
\end{array}
\right.
\]
and $\varepsilon (i)$ is the one-body  DF energy of the state $i$.
The quantities $(vw|K_{12})$ are Clebsch-Gordan coefficients
and $(K_{12}a|K)$  differs from the definition (\ref{CG}) by the
$(-1)^{j_a-m_a}$ factor and $-m_a$ in the $3j$ coefficient.
The indices $v$, $w$, $x$, and $y$ distinguish the
 valence states, indices $a$ and $b$ refer to the
  core states and index $n$ refers to excited states.
Two-particle matrix element $g(1234)$ is the sum of the two-particle
Coulomb and Breit matrix elements and is presented as a product of radial and
angular parts:
\begin{equation}
g(1234)=\sum_{k}\left(X_{k}(1234)+B_k(1234)\right)
(-1)^{j_{1}+j_{2}+m_{1}+m_{2}+k}\times
\sum_{m}(-1)^{m}\left(
\begin{array}{ccc}
j_{1} & j_{3} & k \\
m_{1} & -m_{3} & m
\end{array}
\right) \left(
\begin{array}{ccc}
j_{2} & j_{4} & k \\
m_{2} & -m_{4} & -m
\end{array}
\right).
\end{equation}
Radial part $X_{k}(abcd)$ of the Coulomb matrix element is given by
\begin{equation}
X_{k}(abcd)=\langle a||C_{k}||c\rangle \langle b||C_{k}||d\rangle
R_{k}(abcd)\,.
\end{equation}
The quantities $C_{k}$ are normalized spherical harmonics and
$R_{k}(abcd)$ are Slater integrals.
The expression for the Breit matrix elements $B_k(1234)$ is given
in Ref.~\cite{Breit}. Carrying out angular
reduction, we obtain the final expression for the four contributions $E_{a}$,
$E_{b}$, $E_{c}$, and $E_{d}$ to
second-order diagram of particle-particle-hole interaction:

\begin{multline}\label{eq10}
E_{a}(v^{0}w^{0}[J_{12}]aJ,x^{0}y^{0}[J_{13}]bJ)=\sum_{vw}\sum_{v^{%
\prime }w^{\prime
}}P_{J_{12}}(v^{0}v,w^{0}w)P_{J_{13}}(x^{0}v^{\prime
},y^{0}w^{\prime })\frac{\sqrt{[J_{12}][J_{13}]}}{[J]}(-1)^{J_{12}+J_{13}} \\ 
\times \sum_{n}\sum_{k}\sum_{k^{\prime
}}\frac{X_{k}(vwan)X_{k^{\prime }}(v^{\prime }w^{\prime
}bn))}{\varepsilon (v)+\varepsilon (w)-\varepsilon (a)-\varepsilon
(n)}\delta (J,j_{n})\left\{
\begin{array}{lll}
J_{12} & j_{a} & j_{n} \\
k & j_{w} & j_{v}
\end{array}
\right\} \left\{
\begin{array}{lll}
J_{13} & j_{b} & j_{n} \\
k^{\prime } & j_{w^{\prime }} & j_{v^{\prime }}
\end{array}
\right\} (-1)^{j_{a}+j_{b}+j_{w^{\prime }}+j_{w}+k+k^{\prime }}\,,
\end{multline}
\begin{multline}
E_{b}(v^{0}w^{0}[J_{12}]aJ,x^{0}y^{0}[J_{13}]bJ)=\sum_{vw}\sum_{v^{%
\prime }w^{\prime
}}P_{J_{12}}(v^{0}v,w^{0}w)P_{J_{13}}(x^{0}v^{\prime
},y^{0}w^{\prime })\sqrt{[J_{12}][J_{13}]}   \\ 
\times
\sum_{n}\sum_{k}\sum_{k^{\prime }}\frac{X_{k^{\prime
}}(vww^{\prime }n)}{\varepsilon (v)+\varepsilon (w)-\varepsilon
(w^{\prime })-\varepsilon (n)}\left\{
\begin{array}{lll}
j_{w^{\prime }} & j_{n} & J_{12} \\
j_{w} & j_{v} & k^{\prime }
\end{array}
\right\} (-1)^{1+k+k^{\prime }} \\
\times \left[
+X_{k}(nbav^{\prime })\left\{
\begin{array}{lll}
j_{b} & j_{v^{\prime }} & k \\
j_{w^{\prime }} & J & J_{13}
\end{array}
\right\} \left\{
\begin{array}{lll}
j_{a} & j_{n} & k \\
j_{w^{\prime }} & J & J_{12}
\end{array}
\right\} (-1)^{J_{12}+J_{13}+j_{a}+j_{v^{\prime
}}+j_{n}+j_{w}}\right.  \\
 \left. +X_{k}(nbv^{\prime
}a)\left\{
\begin{array}{lll}
j_{b} & j_{a} & k \\
J_{12} & J_{13} & J
\end{array}
\right\} \left\{
\begin{array}{lll}
j_{n} & j_{v^{\prime }} & k \\
J_{13} & J_{12} & j_{w^{\prime }}
\end{array}
\right\} (-1)^{j_{b}+j_{w^{\prime }}+j_{w}-J}\right]\,,
\end{multline}
\begin{multline}
E_{c}(v^{0}w^{0}[J_{12}]aJ,x^{0}y^{0}[J_{13}]bJ)=\sum_{vw}\sum_{v^{%
\prime }w^{\prime
}}P_{J_{12}}(v^{0}v,w^{0}w)P_{J_{13}}(x^{0}v^{\prime
},y^{0}w^{\prime })\sqrt{[J_{12}][J_{13}]} \\ 
 \times
\sum_{n}\sum_{k}\sum_{k^{\prime }}\frac{X_{k^{\prime }}(v^{\prime
}w^{\prime }vn)}{\varepsilon (v^{\prime })+\varepsilon (w^{\prime
})-\varepsilon (v)-\varepsilon (n)}\left\{
\begin{array}{lll}
j_{v} & j_{n} & J_{13} \\
j_{w^{\prime }} & j_{v^{\prime }} & k^{\prime }
\end{array}
\right\} (-1)^{1+k+k^{\prime }} \\
 \times \left[
X_{k}(wban)\left\{
\begin{array}{lll}
j_{a} & j_{w} & k \\
j_{v} & J & J_{12}
\end{array}
\right\} \left\{
\begin{array}{lll}
j_{b} & j_{n} & k \\
j_{v} & J & J_{13}
\end{array}
\right\} (-1)^{J_{13}+j_{a}+j_{w^{\prime }}+j_{w}+j_{v}}\right.
\\
 \left. +X_{k}(bwan)\left\{
\begin{array}{lll}
j_{b} & j_{a} & k \\
J_{12} & J_{13} & J
\end{array}
\right\} \left\{
\begin{array}{lll}
j_{n} & j_{w} & k \\
J_{12} & J_{13} & j_{v}
\end{array}
\right\} (-1)^{j_{b}+j_{w^{\prime }}+j_{n}+J+J_{12}}\right]\, ,
\end{multline}
\begin{multline}
E_{d}(v^{0}w^{0}[J_{12}]aJ,x^{0}y^{0}[J_{13}]bJ)=\sum_{vw}\sum_{v^{%
\prime }w^{\prime
}}P_{J_{12}}(v^{0}v,w^{0}w)P_{J_{13}}(x^{0}v^{\prime
},y^{0}w^{\prime })\sqrt{[J_{12}][J_{13}]} \\
 \times
\sum_{n}\sum_{kk^{\prime }}\frac{1}{\varepsilon (w)+\varepsilon
(b)-\varepsilon (v^{\prime })-\varepsilon (n)}%
(-1)^{1+j_{n}+j_{w}+j_{a}+j_{v^{\prime }}+J_{12}+J_{13}}
\\
 \times \left[ X_{k}(nvw^{\prime }a)X_{k^{\prime
}}(bwv^{\prime }n)\left\{
\begin{array}{lll}
j_{a} & j_{v} & k \\
j_{w} & J & J_{12}
\end{array}
\right\} \left\{
\begin{array}{lll}
j_{b} & j_{v^{\prime }} & k^{\prime } \\
j_{w^{\prime }} & J & J_{13}
\end{array}
\right\} \left\{
\begin{array}{lll}
j_{w} & j_{n} & k^{\prime } \\
j_{w^{\prime }} & J & k
\end{array}
\right\} \right.  \\
 \left. +X_{k}(nvw^{\prime
}a)X_{k^{\prime }}(wbv^{\prime }n)\left\{
\begin{array}{lll}
j_{a} & j_{v} & k \\
j_{w} & J & J_{12}
\end{array}
\right\} \sum_{\kappa }[\kappa ]\left\{
\begin{array}{lll}
j_{b} & j_{v^{\prime }} & \kappa  \\
j_{w^{\prime }} & J & J_{13}
\end{array}
\right\} \left\{
\begin{array}{lll}
j_{w} & j_{n} & \kappa  \\
j_{w^{\prime }} & J & k
\end{array}
\right\} \left\{
\begin{array}{lll}
j_{b} & j_{v^{\prime }} & \kappa  \\
j_{w} & j_{n} & k^{\prime }
\end{array}
\right\} \right.  \\
 \left. +X_{k}(nvaw^{\prime
})X_{k^{\prime }}(bwv^{\prime }n)\left\{
\begin{array}{lll}
j_{b} & j_{v^{\prime }} & k^{\prime } \\
j_{w^{\prime }} & J & J_{13}
\end{array}
\right\} \sum_{\kappa }[\kappa ]\left\{
\begin{array}{lll}
j_{a} & j_{v} & \kappa  \\
j_{w} & J & J_{12}
\end{array}
\right\} \left\{
\begin{array}{lll}
j_{w^{\prime }} & j_{n} & \kappa  \\
j_{w} & J & k^{\prime }
\end{array}
\right\} \left\{
\begin{array}{lll}
j_{a} & j_{v} & \kappa  \\
j_{w^{\prime }} & j_{n} & k
\end{array}
\right\} \right.  \\
 \left. +X_{k}(nvaw^{\prime
})X_{k^{\prime }}(wbv^{\prime }n)\left\{
\begin{array}{lll}
j_{a} & j_{v} & k \\
j_{w} & J & J_{12}
\end{array}
\right\} \left\{
\begin{array}{lll}
j_{b} & j_{v^{\prime }} & k^{\prime } \\
j_{w^{\prime }} & J & J_{13}
\end{array}
\right\} \sum_{\kappa \kappa ^{\prime }}[\kappa ][\kappa ^{\prime
}]\left\{
\begin{array}{lll}
j_{a} & j_{v} & \kappa  \\
j_{w^{\prime }} & j_{n} & k
\end{array}
\right\} \left\{
\begin{array}{lll}
j_{b} & j_{v^{\prime }} & \kappa ^{\prime } \\
j_{w} & j_{n} & k^{\prime }
\end{array}
\right\} \left\{
\begin{array}{lll}
j_{w} & j_{n} & \kappa ^{\prime } \\
j_{w^{\prime }} & J & \kappa
\end{array}
\right\} \right]\,.
\end{multline}
\end{widetext}


\begin{thebibliography}{28}
\expandafter\ifx\csname natexlab\endcsname\relax\def\natexlab#1{#1}\fi
\expandafter\ifx\csname bibnamefont\endcsname\relax
  \def\bibnamefont#1{#1}\fi
\expandafter\ifx\csname bibfnamefont\endcsname\relax
  \def\bibfnamefont#1{#1}\fi
\expandafter\ifx\csname citenamefont\endcsname\relax
  \def\citenamefont#1{#1}\fi
\expandafter\ifx\csname url\endcsname\relax
  \def\url#1{\texttt{#1}}\fi
\expandafter\ifx\csname urlprefix\endcsname\relax\def\urlprefix{URL }\fi
\providecommand{\bibinfo}[2]{#2}
\providecommand{\eprint}[2][]{\url{#2}}

\bibitem[{\citenamefont{Zhang et~al.}(1989)\citenamefont{Zhang, Sampson, Clark,
  and Mann}}]{m1}
\bibinfo{author}{\bibfnamefont{H.~L.} \bibnamefont{Zhang}},
  \bibinfo{author}{\bibfnamefont{D.~H.} \bibnamefont{Sampson}},
  \bibinfo{author}{\bibfnamefont{R.~E.~H.} \bibnamefont{Clark}},
  \bibnamefont{and} \bibinfo{author}{\bibfnamefont{J.~B.} \bibnamefont{Mann}},
  \bibinfo{journal}{At.\ Data Nucl.\ Data Tables}
  \textbf{\bibinfo{volume}{41}}, \bibinfo{pages}{1} (\bibinfo{year}{1989}).

\bibitem[{\citenamefont{Cowan}(1981)}]{cowan}
\bibinfo{author}{\bibfnamefont{R.~D.} \bibnamefont{Cowan}},
  \emph{\bibinfo{title}{The Theory of Atomic Structure and Spectra}}
  (\bibinfo{publisher}{University of California Press},
  \bibinfo{address}{Berkeley}, \bibinfo{year}{1981}).

\bibitem[{\citenamefont{Chen}(1989)}]{m2}
\bibinfo{author}{\bibfnamefont{M.~H.} \bibnamefont{Chen}},
  \bibinfo{journal}{Phys.\ Rev.\ A} \textbf{\bibinfo{volume}{40}},
  \bibinfo{pages}{2365} (\bibinfo{year}{1989}).

\bibitem[{\citenamefont{Nilsen}(1989)}]{m3}
\bibinfo{author}{\bibfnamefont{J.}~\bibnamefont{Nilsen}},
  \bibinfo{journal}{At.\ Data Nucl.\ Data Tables}
  \textbf{\bibinfo{volume}{41}}, \bibinfo{pages}{131} (\bibinfo{year}{1989}).

\bibitem[{\citenamefont{Bruch et~al.}(1998{\natexlab{a}})\citenamefont{Bruch,
  Safronova, Shlyaptseva, Nilsen, and Schneider}}]{bruch}
\bibinfo{author}{\bibfnamefont{R.}~\bibnamefont{Bruch}},
  \bibinfo{author}{\bibfnamefont{U.~I.} \bibnamefont{Safronova}},
  \bibinfo{author}{\bibfnamefont{A.~S.} \bibnamefont{Shlyaptseva}},
  \bibinfo{author}{\bibfnamefont{J.}~\bibnamefont{Nilsen}}, \bibnamefont{and}
  \bibinfo{author}{\bibfnamefont{D.}~\bibnamefont{Schneider}},
  \bibinfo{journal}{Phys.\ Scr.} \textbf{\bibinfo{volume}{57}},
  \bibinfo{pages}{334} (\bibinfo{year}{1998}{\natexlab{a}}).

\bibitem[{\citenamefont{Bruch et~al.}(1998{\natexlab{b}})\citenamefont{Bruch,
  Safronova, Shlyaptseva, Nilsen, and Schneider}}]{safr}
\bibinfo{author}{\bibfnamefont{R.}~\bibnamefont{Bruch}},
  \bibinfo{author}{\bibfnamefont{U.~I.} \bibnamefont{Safronova}},
  \bibinfo{author}{\bibfnamefont{A.~S.} \bibnamefont{Shlyaptseva}},
  \bibinfo{author}{\bibfnamefont{J.}~\bibnamefont{Nilsen}}, \bibnamefont{and}
  \bibinfo{author}{\bibfnamefont{D.}~\bibnamefont{Schneider}},
  \bibinfo{journal}{J.\ Quant.\ Spectr.\ Radiat.\ Transfer}
  \textbf{\bibinfo{volume}{69}}, \bibinfo{pages}{605}
  (\bibinfo{year}{1998}{\natexlab{b}}).

\bibitem[{\citenamefont{Bautista}(2000)}]{baut}
\bibinfo{author}{\bibfnamefont{M.~A.} \bibnamefont{Bautista}},
  \bibinfo{journal}{J. Phys.\ B} \textbf{\bibinfo{volume}{33}},
  \bibinfo{pages}{71} (\bibinfo{year}{2000}).

\bibitem[{\citenamefont{Eissner et~al.}(1974)\citenamefont{Eissner, Jones, and
  Nussbaumer}}]{eiss74}
\bibinfo{author}{\bibfnamefont{W.}~\bibnamefont{Eissner}},
  \bibinfo{author}{\bibfnamefont{M.}~\bibnamefont{Jones}}, \bibnamefont{and}
  \bibinfo{author}{\bibfnamefont{H.}~\bibnamefont{Nussbaumer}},
  \bibinfo{journal}{Comput.\ Phys.\ Commun.} \textbf{\bibinfo{volume}{8}},
  \bibinfo{pages}{270} (\bibinfo{year}{1974}).

\bibitem[{\citenamefont{Sugar and Corliss}(1985)}]{sugar}
\bibinfo{author}{\bibfnamefont{J.}~\bibnamefont{Sugar}} \bibnamefont{and}
  \bibinfo{author}{\bibfnamefont{C.}~\bibnamefont{Corliss}},
  \bibinfo{journal}{J. Phys.\ Chem.\ Ref.\ Data Suppl.}
  \textbf{\bibinfo{volume}{2}}, \bibinfo{pages}{100} (\bibinfo{year}{1985}).

\bibitem[{\citenamefont{Shirai et~al.}(1990)\citenamefont{Shirai, Funatake,
  Mori, Sugar, Wiese, and Nakai}}]{shirai}
\bibinfo{author}{\bibfnamefont{T.}~\bibnamefont{Shirai}},
  \bibinfo{author}{\bibfnamefont{Y.}~\bibnamefont{Funatake}},
  \bibinfo{author}{\bibfnamefont{K.}~\bibnamefont{Mori}},
  \bibinfo{author}{\bibfnamefont{J.}~\bibnamefont{Sugar}},
  \bibinfo{author}{\bibfnamefont{W.~L.} \bibnamefont{Wiese}}, \bibnamefont{and}
  \bibinfo{author}{\bibfnamefont{Y.}~\bibnamefont{Nakai}}, \bibinfo{journal}{J.
  Phys.\ Chem.\ Ref.\ Data} \textbf{\bibinfo{volume}{19}}, \bibinfo{pages}{127}
  (\bibinfo{year}{1990}).

\bibitem[{\citenamefont{Burkhalter et~al.}(1979)\citenamefont{Burkhalter,
  Cohen, Cowan, and Feldman}}]{a1}
\bibinfo{author}{\bibfnamefont{P.~G.} \bibnamefont{Burkhalter}},
  \bibinfo{author}{\bibfnamefont{L.}~\bibnamefont{Cohen}},
  \bibinfo{author}{\bibfnamefont{R.~D.} \bibnamefont{Cowan}}, \bibnamefont{and}
  \bibinfo{author}{\bibfnamefont{U.}~\bibnamefont{Feldman}},
  \bibinfo{journal}{J.\ Opt.\ Soc.\ Am.} \textbf{\bibinfo{volume}{69}},
  \bibinfo{pages}{1133} (\bibinfo{year}{1979}).

\bibitem[{\citenamefont{Jup{\'{e}}n et~al.}(1988)\citenamefont{Jup{\'{e}}n,
  Engstrom, Hutton, and Tr{\"{a}}abert}}]{a2}
\bibinfo{author}{\bibfnamefont{C.}~\bibnamefont{Jup{\'{e}}n}},
  \bibinfo{author}{\bibfnamefont{L.}~\bibnamefont{Engstrom}},
  \bibinfo{author}{\bibfnamefont{R.}~\bibnamefont{Hutton}}, \bibnamefont{and}
  \bibinfo{author}{\bibfnamefont{E.}~\bibnamefont{Tr{\"{a}}abert}},
  \bibinfo{journal}{J.\ Phys.\ B} \textbf{\bibinfo{volume}{21}},
  \bibinfo{pages}{L347} (\bibinfo{year}{1988}).

\bibitem[{\citenamefont{Schneider et~al.}(1989)\citenamefont{Schneider, Chen,
  Chantrenne, Hutton, and Prior}}]{a3}
\bibinfo{author}{\bibfnamefont{D.}~\bibnamefont{Schneider}},
  \bibinfo{author}{\bibfnamefont{M.~H.} \bibnamefont{Chen}},
  \bibinfo{author}{\bibfnamefont{S.}~\bibnamefont{Chantrenne}},
  \bibinfo{author}{\bibfnamefont{R.}~\bibnamefont{Hutton}}, \bibnamefont{and}
  \bibinfo{author}{\bibfnamefont{M.~H.} \bibnamefont{Prior}},
  \bibinfo{journal}{Phys.\ Rev.\ A} \textbf{\bibinfo{volume}{40}},
  \bibinfo{pages}{4313} (\bibinfo{year}{1989}).

\bibitem[{\citenamefont{Hutton et~al.}(1991)\citenamefont{Hutton, Schneider,
  and Prior}}]{m4}
\bibinfo{author}{\bibfnamefont{R.}~\bibnamefont{Hutton}},
  \bibinfo{author}{\bibfnamefont{D.}~\bibnamefont{Schneider}},
  \bibnamefont{and} \bibinfo{author}{\bibfnamefont{M.~H.} \bibnamefont{Prior}},
  \bibinfo{journal}{Phys.\ Rev.\ A} \textbf{\bibinfo{volume}{44}},
  \bibinfo{pages}{243} (\bibinfo{year}{1991}).

\bibitem[{\citenamefont{Bliman et~al.}(1996)\citenamefont{Bliman, Bruch,
  Altick, Schneider, and Prior}}]{a5}
\bibinfo{author}{\bibfnamefont{S.}~\bibnamefont{Bliman}},
  \bibinfo{author}{\bibfnamefont{R.}~\bibnamefont{Bruch}},
  \bibinfo{author}{\bibfnamefont{P.~L.} \bibnamefont{Altick}},
  \bibinfo{author}{\bibfnamefont{D.}~\bibnamefont{Schneider}},
  \bibnamefont{and} \bibinfo{author}{\bibfnamefont{M.~H.} \bibnamefont{Prior}},
  \bibinfo{journal}{Phys.\ Rev.\ A} \textbf{\bibinfo{volume}{53}},
  \bibinfo{pages}{4176} (\bibinfo{year}{1996}).

\bibitem[{\citenamefont{Phillips et~al.}(1997)\citenamefont{Phillips, Greer,
  Bhatia, Coffey, Barnsley, and Keenan}}]{phil}
\bibinfo{author}{\bibfnamefont{K.~J.~H.} \bibnamefont{Phillips}},
  \bibinfo{author}{\bibfnamefont{C.~J.} \bibnamefont{Greer}},
  \bibinfo{author}{\bibfnamefont{A.~K.} \bibnamefont{Bhatia}},
  \bibinfo{author}{\bibfnamefont{I.~H.} \bibnamefont{Coffey}},
  \bibinfo{author}{\bibfnamefont{R.}~\bibnamefont{Barnsley}}, \bibnamefont{and}
  \bibinfo{author}{\bibfnamefont{F.~P.} \bibnamefont{Keenan}},
  \bibinfo{journal}{Astron.\ Astrophys.} \textbf{\bibinfo{volume}{324}},
  \bibinfo{pages}{381} (\bibinfo{year}{1997}).

\bibitem[{\citenamefont{Brown et~al.}(2001)\citenamefont{Brown, Beiersdorfer,
  Chen, Chen, and Reed}}]{brown}
\bibinfo{author}{\bibfnamefont{G.~V.} \bibnamefont{Brown}},
  \bibinfo{author}{\bibfnamefont{P.}~\bibnamefont{Beiersdorfer}},
  \bibinfo{author}{\bibfnamefont{M.}~\bibnamefont{Chen}},
  \bibinfo{author}{\bibfnamefont{M.~H.} \bibnamefont{Chen}}, \bibnamefont{and}
  \bibinfo{author}{\bibfnamefont{K.}~\bibnamefont{Reed}},
  \bibinfo{journal}{ApJ} \textbf{\bibinfo{volume}{557}}, \bibinfo{pages}{L75}
  (\bibinfo{year}{2001}).

\bibitem[{\citenamefont{Blundell}(1993)}]{qed}
\bibinfo{author}{\bibfnamefont{S.~A.} \bibnamefont{Blundell}},
  \bibinfo{journal}{Phys.\ Rev.\ A} \textbf{\bibinfo{volume}{47}},
  \bibinfo{pages}{1790} (\bibinfo{year}{1993}).

\bibitem[{\citenamefont{Safronova et~al.}(2000)\citenamefont{Safronova,
  Johnson, and Berry}}]{mg}
\bibinfo{author}{\bibfnamefont{U.~I.} \bibnamefont{Safronova}},
  \bibinfo{author}{\bibfnamefont{W.~R.} \bibnamefont{Johnson}},
  \bibnamefont{and} \bibinfo{author}{\bibfnamefont{H.~G.} \bibnamefont{Berry}},
  \bibinfo{journal}{Phys.\ Rev.\ A} \textbf{\bibinfo{volume}{61}},
  \bibinfo{pages}{052503} (\bibinfo{year}{2000}).

\bibitem[{\citenamefont{Safronova et~al.}(2001)\citenamefont{Safronova, Namba,
  Murakami, Johnson, and Safronova}}]{neon}
\bibinfo{author}{\bibfnamefont{U.~I.} \bibnamefont{Safronova}},
  \bibinfo{author}{\bibfnamefont{C.}~\bibnamefont{Namba}},
  \bibinfo{author}{\bibfnamefont{I.}~\bibnamefont{Murakami}},
  \bibinfo{author}{\bibfnamefont{W.~R.} \bibnamefont{Johnson}},
  \bibnamefont{and} \bibinfo{author}{\bibfnamefont{M.~S.}
  \bibnamefont{Safronova}}, \bibinfo{journal}{Phys.\ Rev.\ A}
  \textbf{\bibinfo{volume}{64}}, \bibinfo{pages}{012507}
  (\bibinfo{year}{2001}).

\bibitem[{\citenamefont{Beiersdorfer et~al.}(1986)\citenamefont{Beiersdorfer,
  Bitter, von Goeler, Cohen, Hill, Timberlake, Walling, Chen, Hagelstein, and
  Scofield}}]{b86}
\bibinfo{author}{\bibfnamefont{P.}~\bibnamefont{Beiersdorfer}},
  \bibinfo{author}{\bibfnamefont{M.}~\bibnamefont{Bitter}},
  \bibinfo{author}{\bibfnamefont{S.}~\bibnamefont{von Goeler}},
  \bibinfo{author}{\bibfnamefont{S.}~\bibnamefont{Cohen}},
  \bibinfo{author}{\bibfnamefont{K.~W.} \bibnamefont{Hill}},
  \bibinfo{author}{\bibfnamefont{J.}~\bibnamefont{Timberlake}},
  \bibinfo{author}{\bibfnamefont{R.~S.} \bibnamefont{Walling}},
  \bibinfo{author}{\bibfnamefont{M.~H.} \bibnamefont{Chen}},
  \bibinfo{author}{\bibfnamefont{P.~L.} \bibnamefont{Hagelstein}},
  \bibnamefont{and} \bibinfo{author}{\bibfnamefont{J.~H.}
  \bibnamefont{Scofield}}, \bibinfo{journal}{Phys.\ Rev.\ A}
  \textbf{\bibinfo{volume}{34}}, \bibinfo{pages}{1297} (\bibinfo{year}{1986}).

\bibitem[{\citenamefont{Beiersdorfer et~al.}(1995)\citenamefont{Beiersdorfer,
  Nilsen, Scofield, Bitter, von Goeler, and Hill}}]{b95}
\bibinfo{author}{\bibfnamefont{P.}~\bibnamefont{Beiersdorfer}},
  \bibinfo{author}{\bibfnamefont{J.}~\bibnamefont{Nilsen}},
  \bibinfo{author}{\bibfnamefont{J.~H.} \bibnamefont{Scofield}},
  \bibinfo{author}{\bibfnamefont{M.}~\bibnamefont{Bitter}},
  \bibinfo{author}{\bibfnamefont{S.}~\bibnamefont{von Goeler}},
  \bibnamefont{and} \bibinfo{author}{\bibfnamefont{K.~W.} \bibnamefont{Hill}},
  \bibinfo{journal}{Phys.\ Scr.} \textbf{\bibinfo{volume}{51}},
  \bibinfo{pages}{322} (\bibinfo{year}{1995}).

\bibitem[{\citenamefont{Safronova et~al.}(1996)\citenamefont{Safronova,
  Johnson, and Safronova}}]{bor2}
\bibinfo{author}{\bibfnamefont{M.~S.} \bibnamefont{Safronova}},
  \bibinfo{author}{\bibfnamefont{W.~R.} \bibnamefont{Johnson}},
  \bibnamefont{and} \bibinfo{author}{\bibfnamefont{U.~I.}
  \bibnamefont{Safronova}}, \bibinfo{journal}{Phys.\ Rev.\ A}
  \textbf{\bibinfo{volume}{54}}, \bibinfo{pages}{2850} (\bibinfo{year}{1996}).

\bibitem[{\citenamefont{Safronova et~al.}(1998)\citenamefont{Safronova,
  Johnson, and Safronova}}]{bor3}
\bibinfo{author}{\bibfnamefont{U.~I.} \bibnamefont{Safronova}},
  \bibinfo{author}{\bibfnamefont{W.~R.} \bibnamefont{Johnson}},
  \bibnamefont{and} \bibinfo{author}{\bibfnamefont{M.~S.}
  \bibnamefont{Safronova}}, \bibinfo{journal}{At.\ Data Nucl.\ Data Tables}
  \textbf{\bibinfo{volume}{69}}, \bibinfo{pages}{183} (\bibinfo{year}{1998}).

\bibitem[{\citenamefont{Safronova et~al.}(2002)\citenamefont{Safronova, Namba,
  Albritton, Johnson, and Safronova}}]{alum}
\bibinfo{author}{\bibfnamefont{U.~I.} \bibnamefont{Safronova}},
  \bibinfo{author}{\bibfnamefont{C.}~\bibnamefont{Namba}},
  \bibinfo{author}{\bibfnamefont{J.~R.} \bibnamefont{Albritton}},
  \bibinfo{author}{\bibfnamefont{W.~R.} \bibnamefont{Johnson}},
  \bibnamefont{and} \bibinfo{author}{\bibfnamefont{M.~S.}
  \bibnamefont{Safronova}}, \bibinfo{journal}{Phys.\ Rev.\ A}
  \textbf{\bibinfo{volume}{65}}, \bibinfo{pages}{022507}
  (\bibinfo{year}{2002}).

\bibitem[{\citenamefont{Safronova et~al.}(1997)\citenamefont{Safronova,
  Johnson, and Safronova}}]{be3}
\bibinfo{author}{\bibfnamefont{M.~S.} \bibnamefont{Safronova}},
  \bibinfo{author}{\bibfnamefont{W.~R.} \bibnamefont{Johnson}},
  \bibnamefont{and} \bibinfo{author}{\bibfnamefont{U.~I.}
  \bibnamefont{Safronova}}, \bibinfo{journal}{J. Phys.\ B}
  \textbf{\bibinfo{volume}{30}}, \bibinfo{pages}{2375} (\bibinfo{year}{1997}).

\bibitem[{\citenamefont{Landau and Lifshitz}(1963)}]{Landau}
\bibinfo{author}{\bibfnamefont{L.~D.} \bibnamefont{Landau}} \bibnamefont{and}
  \bibinfo{author}{\bibfnamefont{E.~M.} \bibnamefont{Lifshitz}},
  \emph{\bibinfo{title}{Quantum Mechanics-Non-Relativistic Theory}}
  (\bibinfo{publisher}{Pergamon Press}, \bibinfo{address}{London},
  \bibinfo{year}{1963}).

\bibitem[{\citenamefont{Johnson et~al.}(1988)\citenamefont{Johnson, Blundell,
  and Sapirstein}}]{Breit}
\bibinfo{author}{\bibfnamefont{W.~R.} \bibnamefont{Johnson}},
  \bibinfo{author}{\bibfnamefont{S.~A.} \bibnamefont{Blundell}},
  \bibnamefont{and}
  \bibinfo{author}{\bibfnamefont{J.}~\bibnamefont{Sapirstein}},
  \bibinfo{journal}{Phys.\ Rev.\ A} \textbf{\bibinfo{volume}{37}},
  \bibinfo{pages}{2764} (\bibinfo{year}{1988}).

\end{thebibliography}

\end{document}